\newif{\iffinal}                                
\newif{\iflinks}                                
\newif{\ifarxiv}

\finaltrue{}                                    
\linksfalse{}                                    
\arxivtrue{}

\documentclass[letterpaper,twocolumn,10pt]{article}

\date{}
\title{\Large \bf The Space of Adversarial Strategies}
\author{
    {\rm Ryan Sheatsley*}\\
    University of Wisconsin-Madison
    \and
    {\rm Blaine Hoak*}\\
    University of Wisconsin-Madison
    \and
    {\rm Eric Pauley}\\
    University of Wisconsin-Madison
    \and
    {\rm Patrick McDaniel}\\
    University of Wisconsin-Madison
}

\usepackage{amsmath}                                
\usepackage{balance}                                
\usepackage{booktabs}                               
\usepackage{caption}                                
\usepackage{centernot}                              
\usepackage{colortbl}                               
\usepackage{mathrsfs}                               
\usepackage{float}                                  
\usepackage{graphicx}                               
\usepackage{hyperref}                               
\usepackage{longtable}                              
\usepackage{makecell}                               
\usepackage{mathtools}                              
\usepackage{multirow}                               
\usepackage[all]{nowidow}                           
\usepackage{optidef}                                
\usepackage{pgfplotstable}                          
\usepackage{siis}                                   
\usepackage[binary-units=true,detect-all]{siunitx}  
\usepackage{subcaption}                             
\usepackage{substr}                                 
\usepackage{xfrac}                                  
\usetikzlibrary{chains,shadows.blur,fit,arrows}     

\SetArgSty{textup}                                  
\pgfplotsset{compat=1.16,
    compat/show suggested version=false}            
\providecommand{\qty}{\SI}                          

\newcommand{\aref}[1]{\hyperref[#1]{appendix~\ref*{#1}}}
\makeatletter\newcommand\notsotiny{\@setfontsize\notsotiny{6}{7}}\makeatother


\newcommand{\hypertargetblue}[2]{\textbf{\textcolor{blue}{\hypertarget{#1}{#2}}}}

\newcommand{\pytorch}{\texttt{PyTorch}}             
\newcommand{\sgdfull}{
    \texttt{Stochastic Gradient Descent}}
\newcommand{\sgd}{\texttt{SGD}}                     
\newcommand{\adam}{\texttt{Adam}}                   
\newcommand{\mbsfull}{
    \texttt{Momentum Best Start}}
\newcommand{\mbs}{\texttt{MBS}}                     
\newcommand{\bwsgdfull}{
    \texttt{Backward Stochastic Gradient Descent}}
\newcommand{\bwsgd}{\texttt{BWSGD}}                 
\newcommand{\crossentropyfull}{
    \texttt{Cross-Entropy}}
\newcommand{\crossentropy}{\texttt{CE}}             
\newcommand{\identitylossfull}{
    \texttt{Identity Loss}}
\newcommand{\identityloss}{\texttt{IL}}             
\newcommand{\cwlossfull}{
    \texttt{Carlini-Wagner Loss}}
\newcommand{\cwloss}{\texttt{CWL}}                  
\newcommand{\dlrlossfull}{
    \texttt{Difference of Logits Ratio Loss}}
\newcommand{\dlrloss}{\texttt{DLR}}                 

\newcommand{\rr}{\textit{Random-Restart}}           
\newcommand{\cov}{\textit{Change of Variables}}     
\newcommand{\fgsmfull}{\texttt{
    Fast Gradient Sign Method}}
\newcommand{\fgsm}{\texttt{FGSM}}                   
\newcommand{\bimfull}{\texttt{
    Basic Iterative Method}}
\newcommand{\bim}{\texttt{BIM}}                     
\newcommand{\pgdfull}{\texttt{
    Projected Gradient Descent}}
\newcommand{\pgd}{\texttt{PGD}}                     
\newcommand{\jsmafull}{\texttt{
    Jacobian-based Saliency Map Approach}}
\newcommand{\jsma}{\texttt{JSMA}}                   
\newcommand{\dffull}{\texttt{
    DeepFool}}
\newcommand{\df}{\texttt{DF}}                       
\newcommand{\cwfull}{\texttt{
    Carlini-Wagner Attack}}
\newcommand{\cw}{\texttt{CW}}                       
\newcommand{\autoattackfull}{\texttt{
    AutoAttack}}
\newcommand{\autoattack}{\texttt{AA}}               

\newcommand{\apgdcefull}{\texttt{
    Auto Projected Gradient Descent - Cross Entropy}}
\newcommand{\apgdce}{\texttt{APGD-CE}}              
\newcommand{\apgddlrfull}{\texttt{
    Auto Projected Gradient Descent - Difference of Logits Ratio}}
\newcommand{\apgddlr}{\texttt{APGD-DLR}}                     
\newcommand{\fabfull}{\texttt{
    Fast Adaptive Boundary Attack}}
\newcommand{\fab}{\texttt{FAB}}                     
\newcommand{\fw}[1][]{%
    \texttt{ATK\textsubscript{\texttt{#1}}}}        
\newcommand{\oeafull}{\texttt{
    Pareto Ensemble Attack}}
\newcommand{\oea}{\texttt{PEA}}                     
\newcommand{\jsmasmapfull}{
    \texttt{Jacobian Saliency Map}}
\newcommand{\jsmasmap}{%
    \texttt{SM\textsubscript{J}}}                   
\newcommand{\dfsmapfull}{
    \texttt{DeepFool Saliency Map}}
\newcommand{\dfsmap}{
    \texttt{SM\textsubscript{D}}}
\newcommand{\identitysmapfull}{
    \texttt{Identity Saliency Map}}
\newcommand{\identitysmap}{
    \texttt{SM\textsubscript{I}}}
\newcommand{\phishing}{\texttt{Phishing}}           
\newcommand{\nslkdd}{\texttt{NSL-KDD}}              
\newcommand{\unswnb}{\texttt{UNSW-NB15}}            
\newcommand{\mnist}{\texttt{MNIST}}                 
\newcommand{\fmnistfull}{
    \texttt{Fashion-MNIST}}
\newcommand{\fmnist}{\texttt{FMNIST}}               
\newcommand{\cifar}{\texttt{CIFAR-10}}              
\newcommand{\malmemfull}{\texttt{CIC-MalMem-2022}}  
\newcommand{\malmem}{\texttt{MalMem}}               

\newcommand{\mr}[1]{[MR.#1]}                        

\usepackage{usenix-2020-09}

\begin{document}
\maketitle
\iffinal{}
\begingroup\renewcommand\thefootnote{*}
\footnotetext{Equal contribution}
\endgroup
\fi{}
\begin{abstract}\label{abstract}

    \noindent \textit{Adversarial examples}, inputs designed to induce worst-case
    behavior in machine learning models, have been extensively studied over the
    past decade. Yet, our understanding of this phenomenon stems from a rather
    fragmented pool of knowledge; at present, there are a handful of attacks,
    each with disparate assumptions in threat models and incomparable
    definitions of optimality. In this paper, we propose a systematic approach
    to characterize worst-case (\ie{} optimal) adversaries. We first introduce
    an extensible decomposition of attacks in adversarial machine learning by
    atomizing attack components into \textit{surfaces} and \textit{travelers}.
    With our decomposition, we enumerate over components to create \num{576}
    attacks (\num{568} of which were previously unexplored). Next, we propose
    the \oeafull{} (\oea{}): a theoretical attack that upper-bounds attack
    performance. With our new attacks, we measure performance relative to the
    \oea{} on: both robust and non-robust models, seven datasets, and three
    extended \lp-based threat models incorporating compute costs, formalizing
    the \textit{Space of Adversarial Strategies}. From our evaluation we find
    that attack performance to be highly contextual: the domain, model
    robustness, and threat model can have a profound influence on attack
    efficacy.
    Our investigation suggests that future studies measuring the security of
    machine learning should: (1) be contextualized to the domain \&
    threat models, and (2) go beyond the handful of known attacks used today.

\end{abstract}

\thispagestyle{plain}
\pagestyle{plain}
\section{Introduction}\label{introduction}

It is well-known that machine learning models are vulnerable to
\textit{adversarial examples}---inputs designed to induce worst-case behavior.
Seminal papers have introduced a suite of varying techniques for producing
adversarial examples, each with their own unique threat models, strengths, and
weaknesses~\cite{goodfellow_explaining_2014, papernot_limitations_2016,
carlini_towards_2017, moosavi-dezfooli_deepfool_2016, madry_towards_2017}.
Every generation of research yields the next evolution of attacks, designed to
overcome prior defenses. It is unclear whether this evolution will ever
converge, yet it is apparent that there are some attacks that have ``survived''
modern defenses. Specifically, the accepted baselines for evaluating defenses
are converging to a small set of largely fixed attacks and threat models.

This observation on the fixed nature of commonly used attacks speaks to a
broader and more fundamental problem in the way we evaluate the trustworthiness
of machine learning systems: our understanding of adversaries has been derived
from a union of works with disjoint assumptions and underlying threat models.
As a consequence, it is challenging to draw any universal truths from a rather
fragmented (and broadly incomparable) pool of knowledge. Subsequently,
comparisons between attacks and attempts at characterizing \textit{the}
worst-case adversary have been through the lens of a specific threat model and
defined with respect to a small handful of attacks, making it difficult to
discern the true strength of claims on what is good (or even best) and when.


In this paper, we introduce a systematic approach to determine worst-case
adversaries. We first introduce \num{568} new attacks by anatomizing seminal
attacks into interchangeable components, therein enabling a meaningful
evaluation of model robustness against an expansive attack space. With this
decomposition, we formalize an extensible \textit{Space of Adversarial
Strategies}: the set of attacks considered by an adversary under a specific
threat model and domain. We then empirically approximate the \oeafull{}
(\oea{}): a theoretical attack which upper-bounds attack performance by
returning the optimal set of adversarial examples for a given threat model and
dataset. We then use the \oea{} to explore a fundamental question: \textit{Does
an optimal attack exist?}

Our analysis begins by decomposing seminal attacks in adversarial machine
learning. We observe that all known attacks are broadly built from two
components: (1) a \textit{surface}, and (2) a \textit{traveler}. Surfaces
encode the traversable attack space (often as the gradient of a
cost function), while travelers are ``vehicles'' that navigate a surface to
meet adversarial goals. Attack components live within surfaces and travelers,
which characterize attack behavior, such as building crude surfaces that favor
meeting adversarial goals without regard to budget, or vice-versa. Our
decomposition allows us to (a) generalize attacks in an extensible manner, and
(b) naturally construct new (and known) attacks by permuting attack components.

From our decomposition, we permute attack components to build a previously
unexplored attack space, yielding \num{568} new attacks. We then measure attack
performance through the \oea{}, which is built by forming the \textit{lower
envelope} of measured model accuracy across attacks over the budget consumed.
In other words, the \oea{} bounds the performance an individual attack could
achieve. We rank attacks with respect to the \oea{} by measuring the difference
in areas of their performance curves. Our approach not only gives us a
comparable definition of optimality, but also a mechanism by which we can
measure the merit of individual attacks.

\iflinks\hypertargetblue{mr3}{\mr{3}}\fi Our evaluation across seven datasets, three threat models, and robust (through
adversarial training) versus non-robust models found relative attack performance to be
highly contextual. Specifically, (1) the domain and threat model can have a profound
effect (especially if the trained model is robust), and
(2) even the advantage of certain component choices is sensitive to these factors, as well as other paired components.

\noindent We make the following contributions:
\vspace{-2mm}
\begin{itemize}

    \item We propose a decomposition of attacks in adversarial machine learning
        by atomizing attack components into two main layers, \textit{surfaces}
        and \textit{travelers}. Our decomposition readily enables extensions of
        new components.

    \item We characterize the attack space by permuting components of known
        attacks, yielding \num{568} new attacks.

    \item We introduce a systematic approach to compare the efficacy of
        attacks. We first build the \oeafull{} from the performance curves of
        attacks and rank their relative performance.

    \item We instantiate and enumerate over a hypothesis space to identify
        which strategies perform better than others under a given threat model.


\end{itemize}

\section{Background}\label{background}


\subsection{Threat Models}\label{background:threats}

Adversaries have historically had one of two goals: minimizing model
accuracy~\cite{papernot_limitations_2016, moosavi-dezfooli_deepfool_2016, lowd_adversarial_2005, szegedy_intriguing_2013} or
maximizing model loss~\cite{goodfellow_explaining_2014, madry_towards_2017, biggio_evasion_2013}.
The risks associated with minimizing model accuracy are often exemplified by
vehicles misclassifying traffic signs~\cite{eykholt_robust_2018}, intrusion
detection systems permitting malicious entities
entry~\cite{yang_adversarial_2018}, medical
misdiagnoses~\cite{finlayson_adversarial_2019}, among other failures.
Maximizing model loss serves two purposes: (1) it is a surrogate for minimizing
model accuracy (as, the inverse is performed to maximize model accuracy during
model training), and (2) it aids in \iflinks\hypertargetblue{mr6b}{\mr{6b}}\fi transferability
attacks~\cite{papernot_transferability_2016, tramer_space_2017,
papernot_practical_2017, demontis_why_2019}. In this work, we focus on
minimizing model accuracy and defer the explorations of transferability to
future work.

\iflinks\hypertargetblue{mr4a}{\mr{4a}}\fi In the context of minimizing model accuracy, translating the risks above into
an optimization objective to be solved by an adversary is commonly written as:
\begin{argmini}|l|
    {\epsilon}{\norm{\epsilon}_{p}}{}{}
    \addConstraint{f (x + \epsilon) \neq{} \hat{y}}
    \addConstraint{x + \epsilon\in\ball{\phi}{x}}.%
    \label{eq:goals}
\end{argmini}

\noindent where we are given a victim model \(f\), a sample \(x\), label
\(\hat{y}\), a self-imposed budget \(\phi\) measured under some \lp{}-norm.
Conceptually, the adversary searches within some self-imposed norm-ball
\(\mathcal{B}\) of radius \(\phi\), centered at \(x\) for a ``small'' change
\(\epsilon\) that, when applied to \(x\), yields the desired goal. 

With adversarial goals and capabilities defined, the final component of threat
models pertains to \textit{access}. Specifically, subsequent works have shown
that adversaries need not have direct access to the victim model \(f\) to
produce adversarial examples; models trained on similar data have similar
decision manifolds, and thus, adversarial examples can ``transfer'' from one
model to another~\cite{papernot_transferability_2016, tramer_space_2017,
papernot_practical_2017}. When access is restricted (and thus, transferability
is exploited), such threat models are called ``grey-'' or ``black-box'', while
full access to the victim model is called a ``white-box'' threat model. In this
paper, we focus on white-box threat models as they represent the worst-case
adversaries (in that they can produce adversarial examples with the tightest
\lp{}-norm constraints). However, our decomposition and performance
measurements can be directly applied to grey- and black-box threat models as
well, which we further discuss in \autoref{discussion}.

\shortsection{On \lp{}-norms}\label{background:norms} As shown in \autoref{eq:goals}, the ``cost'' for crafting adversarial examples
has been predominantly measured through \lp{}-norms. Informally, adversarial
examples induce a misclassification between human and machine; \lp{}-bounded
examples attempt to meet this definition. This concept arose from attacks on
images, in that attacks would produce adversarial examples whose perturbations
were invisible to humans, yet influential on models. \lp{}-norms are
becoming an increasingly controversial topic, in that it has been debated if
they have meaningful interpretations in non-visual
domains~\cite{sheatsley_robustness_2021}, or even visual
domains~\cite{chen_explore_2020}, or if they are useful at
all~\cite{sharif_suitability_2018}. Regardless, attacks have broadly converged
on optimizing under \lp[0], \lp[2], or \lp[\infty], and thus we focus our study
on those.

\subsection{Attack Algorithms}\label{background:attacks}

Here we briefly discuss the attack algorithms used in our decomposition
(specifically, the unique components they introduce). We study these algorithms
specifically due to their prevalence across works in adversarial machine
learning~\cite{ren_adversarial_2020}.

\shortsection{\bimfull{} (\bim{})} \bim{}~\cite{kurakin_adversarial_2016} is an
iterative extension of \fgsmfull{} (\fgsm{})~\cite{goodfellow_explaining_2014}.
\bim{} is an \lp[\infty]-based attack that perturbs based on the gradient of a
cost function, typically \crossentropyfull{} (\crossentropy{}). It often uses
\sgdfull{} (\sgd{}) as its optimizer for finding adversarial examples.

\shortsection{\pgdfull{} (\pgd{})} \pgd{}~\cite{madry_towards_2017} is widely
regarded as the state-of-the-art in crafting algorithms. \pgd{} is identical to
\bim{}, with the exception of a \rr{} preprocessing step, wherein inputs are
initially randomly perturbed within an \lp[\infty] ball.

\shortsection{\jsmafull{} (\jsma{})} The
\jsma{}~\cite{papernot_limitations_2016} is an \lp[0]-based attack that is
unique in its definition of a \textit{saliency map}; a heuristic applied to the
model Jacobian to determine the most salient feature to perturb in a given
iteration. Unlike most other attacks, it does not rely on a cost function, but
rather uses the model Jacobian directly. In our decomposition, we denote the
\jsma{} saliency map as \jsmasmap{}. The \jsma{} uses \sgd{} as its optimizer.

\shortsection{\dffull{} (\df{})} \df{}~\cite{moosavi-dezfooli_deepfool_2016} is
an \lp[2]-based attack which models crafting adversarial examples as a
projection onto the decision boundary. We find that we can model this
projection as a saliency map, much like the \jsma{}, which we denote as
\dfsmap{}. Similar to the \jsma{}, \df{} relies on the model Jacobian, does not
have a cost function, and uses \sgd{}.

\shortsection{\cwfull{} (\cw{})} \cw{}~\cite{carlini_towards_2017} is an
\lp[2]-based attack that is unique across several dimensions: (1) it uses a
custom loss, which we label \cwlossfull{} (\cwloss{}), (2) introduces the
\cov{} technique, which ensures that, during crafting, the intermediate
adversarial examples always comply with a set of box constraints, and (3) uses
\adam{} as its optimizer for finding adversarial examples.

\shortsection{\autoattackfull{} (\autoattack{})}
\autoattack{}~\cite{croce_reliable_2020} is an ensemble attack consisting of
three different white-box attacks (as well as one black-box attack). This
ensemble is unique in that all of its attacks are parameter free (except for
the number of iterations to run attacks for). Its white-box attacks are: (1)
\apgdcefull{} (\apgdce{}), which is \pgd{} with the \mbsfull{} optimizer, (2)
\apgddlrfull{} (\apgddlr{}), which is \apgdce{} but with \dlrlossfull{}, and
(3) \fabfull{}~\cite{croce_minimally_2020} (\fab{}), which is similar to
\dffull{}, but it's optimizer \bwsgdfull{} applies a biased gradient step and a
backward step to stay close to the original point.

\section{Decomposing AML}\label{decomposition}

\iflinks\hypertargetblue{mr1}{\mr{1}}\fi From analysis of popular attacks
(discussed in \autoref{background:attacks}), we find that attacks broadly
perform two main functions to produce adversarial examples, they: (1)
manipulate \(x\), such as with \rr{}, or (2) manipulate gradients, such as by
using a saliency map. We use this observation as a starting point for our
decomposition; components that do the former are part of the \textit{traveler}
and ones that do the latter are within the \textit{surface}. Through this
generalization, an attack can be seen as, simply, a choice of values for each
of these components rather than a unique, incomparable entity.


Importantly, these components are broadly \textit{mutually compatible} with one
another, in that one could omit, add, or swap them when building an attack. We
exploit this property when permuting components, therein yielding a vast space
of attacks, some of which are known, but most of which are not. This modular view of attacks not only allows us to build this vast space, but also makes the framework highly extensible by nature; new attacks can add on new choices for components or even new components entirely. A summary of
the evaluated components in this paper and the compositions of well-known
attacks are shown in \autoref{tab:components}.

\begin{figure}[t]
    \centering
    \resizebox{.49\textwidth}{!}{\begin{tikzpicture}[auto,
  node distance = 10mm,
  start chain = going below,
  box/.style = {draw,rounded corners,blur shadow,fill=white,very thick,
        on chain,align=center}]
    \node[] (travelername) {\textbf{Traveler}};
    \node[box,rounded corners,thin,dotted, below = 2mm of travelername,fill=r!30] (RR)    {$x \leftarrow$ \rr($x$)}; 
    \node[box,rounded corners,thin,dotted,fill=orange!30] (CoV) {$x \leftarrow$ \cov($x$)};
    \node[box,fill=ye!30] (optim) {\textit{Optimizer}($x$, $\alpha$)};
    \node[box] (update) {$x_{i+1} \leftarrow x_i + \alpha \cdot \nabla_{x_i}$\textit{Surface}};
    \node[fit=(RR) (CoV) (optim) (update),draw,minimum width=4.5cm,minimum height=5.5cm,dashed] (traveler) {};
    
    \node[box, right = 30mm of update,fill=green!30] (norm) {$\nabla_{x_i}$\textit{Surface} $\leftarrow \lp (SM$, $p)$ };
    \node[box,rounded corners, thin,dotted,above = 10mm of norm,fill=blue!30] (salmap) {$SM \leftarrow$ \textit{SaliencyMap} ($\nabla_{x_i} Loss$) };
    \node[box,rounded corners, thin,dotted,above = 8mm of salmap,fill=iris!30] (loss) {$\nabla_{x_i} Loss \leftarrow  \frac{\partial Loss}{\partial f(x_i)} \cdot J $};
    \node[box, above = 8mm of loss,fill=violet!30] (jac) {$J \leftarrow \frac{\partial f(x_i)}{\partial x_i} $};
    \node[above = 2mm of jac] (surfacename) {\textbf{Surface}};
    \node[fit=(jac) (loss) (salmap) (norm),draw,minimum width=4.5cm,minimum height=5.5cm,dashed] (surface) {};
 \begin{scope}[rounded corners,-latex]
  \path (RR) edge (CoV) (CoV) edge (optim) (optim) edge (update) (jac) edge (loss) (loss) edge (salmap) (salmap) edge (norm);
  \draw ([xshift=12mm]update.45) -- ++ (0,2) -| node[above,pos=.25]{$x_i$} ([xshift=-10mm]jac.west) -- (jac.west);
 \path ([yshift=-6mm]norm) edge node[above]{$\nabla_{x_i}$\textit{Surface}} (update);
 \end{scope}
\end{tikzpicture}}
    \caption{Flow of composition between the surface and traveler to construct
    an attack. Required components have bold outlines while optional components
    have dotted outlines.}\label{fig:decomp_flow}
\end{figure}
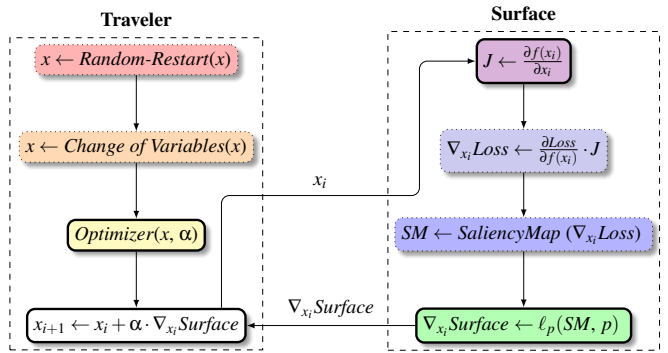

For the remainder of this section, we  describe: (1) the components that
constitute a surface and their options, (2) the layers that define a traveler
and associated configurations, and (3) a characterization of the attack space.
An overview of the composition of the surface and traveler, and their
interaction is shown in \autoref{fig:decomp_flow}. All symbols defined in this
section (and in the remainder of the paper) can be found in \aref{appendix-e}.

\begin{table}
    \resizebox{\columnwidth}{!}{%
    \begin{tabular}{rl rl}
        \toprule
        \multicolumn{4}{c}{\textbf{Attack Algorithms}}\\
        \cmidrule(l{4.2em}r{3.6em}){1-4}
        \multicolumn{2}{c}{\textbf{Surface Components}}&
        \multicolumn{2}{c}{\textbf{Traveler Components}}\\
        \midrule
        \textit{Losses:} &\makecell[tl]{\crossentropyfull{}\\\cwlossfull{}\\
            \identitylossfull{}\\\dlrlossfull{}}
        & \rr{}: & Enabled, Disabled\\
        \textit{Saliency Maps:} & \jsmasmap{}, \dfsmap{}, \identitysmap{}
        & \cov{}: & Enabled, Disabled\\
        \textit{\lp{}-norm} & \lp[0], \lp[2], \lp[\infty]
        & \textit{Optimizer:} & \sgd{}, \adam{}, \mbs{}, \bwsgd{} \\
        \bottomrule
    \end{tabular}
    }
    \rule{0pt}{8ex}
    \resizebox{\columnwidth}{!}{%
    \begin{tabular}{rcccccccccccccccc}
        \bim{} &                                                
        \colcir[bl] & \colcir[wh] & \colcir[wh] & \colcir[wh] & 
        \colcir[wh] & \colcir[wh] & \colcir[bl] &               
        \colcir[wh] & \colcir[wh] & \colcir[bl] &               
        \colcir[wh] &                                           
        \colcir[wh] &                                           
        \colcir[bl] & \colcir[wh] & \colcir[wh] & \colcir[wh]\\ 

        \pgd{} &
        \colcir[bl] & \colcir[wh] & \colcir[wh] &  \colcir[wh] &
        \colcir[wh] & \colcir[wh] & \colcir[bl] &
        \colcir[wh] & \colcir[wh] & \colcir[bl] &
        \colcir[bl] &
        \colcir[wh] &
        \colcir[bl] & \colcir[wh] & \colcir[wh] & \colcir[wh] \\

        \jsma{} &
        \colcir[wh] & \colcir[wh] & \colcir[bl] & \colcir[wh]&
        \colcir[bl] & \colcir[wh] & \colcir[wh] &
        \colcir[bl] & \colcir[wh] & \colcir[wh] &
        \colcir[wh] &
        \colcir[wh] &
        \colcir[bl] & \colcir[wh] & \colcir[wh]& \colcir[wh]\\

        \df{} &
        \colcir[wh] & \colcir[wh] & \colcir[bl] & \colcir[wh]&
        \colcir[wh] & \colcir[bl] & \colcir[wh] &
        \colcir[wh] & \colcir[bl] & \colcir[wh] &
        \colcir[wh] &
        \colcir[wh] &
        \colcir[bl] & \colcir[wh] & \colcir[wh]& \colcir[wh]\\

        \cw{} &
        \colcir[wh] & \colcir[bl] & \colcir[wh] & \colcir[wh]&
        \colcir[wh] & \colcir[wh] & \colcir[bl] &
        \colcir[wh] & \colcir[bl] & \colcir[wh] &
        \colcir[wh] &
        \colcir[bl] &
        \colcir[wh] & \colcir[bl] & \colcir[wh]& \colcir[wh]\\

        \apgdce{} &
        \colcir[bl] & \colcir[wh] & \colcir[wh] &  \colcir[wh] &
        \colcir[wh] & \colcir[wh] & \colcir[bl] &
        \colcir[wh] & \colcir[wh] & \colcir[bl] &
        \colcir[bl] &
        \colcir[wh] &
        \colcir[wh] & \colcir[wh] & \colcir[bl] & \colcir[wh] \\

        \apgddlr{} &
        \colcir[wh] & \colcir[wh] & \colcir[wh] &  \colcir[bl] &
        \colcir[wh] & \colcir[wh] & \colcir[bl] &
        \colcir[wh] & \colcir[wh] & \colcir[bl] &
        \colcir[bl] &
        \colcir[wh] &
        \colcir[wh] & \colcir[wh] & \colcir[bl] & \colcir[wh] \\

        \fab{} &
        \colcir[wh] & \colcir[wh] & \colcir[bl] & \colcir[wh]&
        \colcir[wh] & \colcir[bl] & \colcir[wh] &
        \colcir[wh] & \colcir[bl] & \colcir[wh] &
        \colcir[wh] &
        \colcir[wh] &
        \colcir[wh] & \colcir[wh] & \colcir[wh]& \colcir[bl]\\
        \midrule
        & \rotatebox[origin=c]{-45}{\crossentropy{}} &
        \rotatebox[origin=c]{-45}{\cwloss{}} &
        \rotatebox[origin=c]{-45}{\identityloss{}} &
        \rotatebox[origin=c]{-45}{\dlrloss{}} &
        \rotatebox[origin=c]{-45}{\jsmasmap{}} &
        \rotatebox[origin=c]{-45}{\dfsmap{}} &
        \rotatebox[origin=c]{-45}{\identitysmap{}} &
        \rotatebox[origin=c]{-45}{\lp[0]} &
        \rotatebox[origin=c]{-45}{\lp[2]} &
        \rotatebox[origin=c]{-45}{\lp[\infty]} &
        \rotatebox[origin=c]{-45}{RR} &
        \rotatebox[origin=c]{-45}{CoV} &
        \rotatebox[origin=c]{-45}{\sgd{}} &
        \rotatebox[origin=c]{-45}{\adam{}} &
        \rotatebox[origin=c]{-45}{\mbs{}} &
        \rotatebox[origin=c]{-45}{\bwsgd{}} \\
        \bottomrule
    \end{tabular}
    }
    \caption{Attack Component Decomposition.}\label{tab:components}
\end{table}

\subsection{Surfaces}

Surfaces, which encode the traversable attack space, are built from: (1) the
model Jacobian, (2) the gradient of a loss function, (3) the application of a
saliency map, and (4) an \lp{}-norm. 
context of crafting adversarial examples.

\shortsection{Model Jacobian} At the heart of every surface (and thus, every
attack) is the model Jacobian. The Jacobian \(\jacobian{}\) of a model with
respect to a sample \(x\) encodes the influence each feature \(i\) in \(x\) has
over each class. While most attack papers encode perturbations as a function of
the gradient of a loss function, such computations \textit{necessarily} involve
computing a portion (at least) of the model Jacobian (whether attacks require
the full model Jacobian is a matter of design choice). This is evident via
application of the chain rule:
\begin{equation*}
    \frac{\partial L(f(x), \hat{y})}{\partial x} = \frac{%
        \partial L(f(x),\hat{y})}{\partial f(x)}\cdot \frac{%
            \partial f(x)}{\partial x} = \frac{\partial L(f(x),\hat{y})}{\partial f(x)}\cdot \textbf{J}
\end{equation*}

\noindent Importantly, computing a Jacobian is computationally expensive, on
the order of \(\bigo{}(d\cdot c)\), where \(d\) describes the dimensionality of
\(x\) (\ie{} the number of features) and \(c\) describes the number of classes.
Thus, attacks that \textit{require} the full model Jacobian (\eg{} \jsma{} and
\df{}) must pay a (sometimes substantial) cost in compute resources to produce
adversarial examples---a fact largely overlooked. This component is perhaps the one with the greatest potential for extensibility. For instance, black-box attacks or those wanting to overcome obfuscated gradients~\cite{athalye_obfuscated_2018} could opt to use Backwards Pass Differentiable Approximation (BPDA)~\cite{athalye_obfuscated_2018} to obtain a jacobian rather than a traditional backwards pass. 

\shortsection{Loss Functions} Perhaps \textit{the} most popular design choice
in attack algorithms is to perturb features based on the gradient of a loss
function. The intuition is straightforward: we rely on surrogate measurements
to learn parameters that have maximal accuracy during training, and thus, we
can exploit these same measures to produce samples that induce minimal
accuracy. This is commonly \crossentropyfull{} (\crossentropy{}) loss:
\vspace{-1.5mm}
\begin{equation*}
    \sum_{i}^c -\hat{y}_i\cdot \log(y_i)
\end{equation*}

\vspace{-1.5mm}
\noindent where \(c\) is the number of classes, \(\hat{y}_i\) is the label as a
one-hot encoded vector, and \(y_i\) is the output of the \texttt{softmax}
function.

Aside from \crossentropy{} loss, other attack philosophies instead opt for
custom loss functions that explicitly encode adversarial objectives, such as
\cwlossfull{} (\cwloss{}):
\vspace{-1.5mm}
\begin{equation*}
    \norm{\delta}^p_p + c\cdot\max(f_{\hat{y}}(x)-\max\{f_{i}(x) : i \neq \hat{y} \},0)
\end{equation*}

\noindent where \(p\) is the target \lp{}-norm to optimize under and \(c\) is a
hyperparameter that controls the trade-off between the distortion introduced
and misclassification.

Similar to the latter half of \cwloss{}, the \dlrlossfull{} (\dlrloss{}) takes
the difference between the true logit and the largest non-true-class logit.
However, this loss function also divides by the difference between the largest
logit (\(f_{\pi_1}(x)\)) and the third largest logit (\(f_{\pi_3}(x)\)), as
follows:
\begin{equation*}
    -\frac{f_{\hat{y}}(x)-\max\{f_{i}(x) : i \neq \hat{y}\}}{f_{\pi_1}(x) - f_{\pi_3}(x)}
\end{equation*}

Finally, some attacks do not have an explicit loss function (such as \jsma{} or
\df{}) and instead rely on information at other layers in the surface to
produce adversarial examples (\eg{} through saliency maps). To support this
generalization, we implement a pseudo-identity loss function,
\identitylossfull{} (\identityloss{}), which simply returns the \(\hat{y}\)th
model logit component.

\shortsection{Saliency Maps} Saliency maps, in the context of adversarial
machine learning, were first introduced by the
\jsma{}~\cite{papernot_limitations_2016}. These maps encode heuristics to best
achieve adversarial goals by coalescing model Jacobian information into a
gradient. We slightly tweak the original definition of the saliency map used in
the \jsma{} to be: (1) independent
of perturbation direction, and (2) agnostic of a target class. Though
functionally different, we call this saliency map the \jsmasmapfull{}
(\jsmasmap{}), as the underlying heuristic is identical in spirit to the one
introduced by the \jsma{}:
\begin{equation*}
    {\jsmasmap{}}_i\left(\hat{y},\jacobian\right) =
    \begin{dcases}
        0 & \textrm{if } \sign(J_{\hat{y},i}) = \sign(\sum_{j \neq \hat{y}} J_{j,i})\\
        \abs{J_{\hat{y},i}} \cdot \sum_{j \neq \hat{y}} J_{j,i} & \text{otherwise}
    \end{dcases}
\end{equation*}

\noindent where \(\hat{y}\) is the label for a sample \(x\), \(\jacobian\) is
the Jacobian of a model with respect to \(x\), and \(i\) is the \(i\)th feature
of \(x\). Moreover, we observe that attack formulations with complex heuristics, such as \dffull{}, can be cast as-is into a saliency
map as well. We define the \dfsmapfull{} (\dfsmap{}) as:
\begin{equation*}
    {\dfsmap{}}\left(x, \hat{y}, q\right) =
    \frac{\abs{f_{\hat{y}}(x)-f_{k}(x)}}{%
        \norm{J_{\hat{y}}-J_{k}}_{q}^{q}}
        \cdot{(J_{\hat{y}}-J_{k})}^{q-1} \cdot\sign(J_{\hat{y}}-J_{k})
\end{equation*}

\noindent where \(x\) is a sample, \(\hat{y}\) is the label for \(x\), \(q\) is
calculated from the \lp{} norm where $q=\frac{p}{p-1}$,  \(f\) is the model,
and \(k\) is the ``closest'' class to the true label \(\hat{y}\) calculated by:
\begin{equation*}
    k = \argmin_{i \neq \hat{y}}
    \frac{\abs{f_{\hat{y}}(x)-f_{i}(x)}}{%
        \norm{J_{\hat{y}}-J_{i}}_{q}}
\end{equation*}

\noindent Notably, unlike the \jsmasmap{}, this formulation is identical to
that presented in the original \dffull{} attack. 

\iflinks\hypertargetblue{mr4b}{\mr{4b}}\fi Finally, attacks can also opt
not to use any form of saliency map, and thus, we define an identity saliency
map, \identitysmapfull{} (\identitysmap{}), which simply returns the passed-in
gradient-like information as-is.

\shortsection{\lp{}-norms} To meet threat model constraints, nearly all attacks
manipulate gradient information via an \lp{}-norm. We remark that this can be
conceptualized as a layer in a surface. Thus, we provide abstractions for three
popular \lp{}-based threat models, defined as:
\begin{equation*}
    \begin{aligned}
        \ell_\infty(\nabla) &= \sign(\nabla)\\
        \ell_2(\nabla) &= \frac{\nabla}{\norm{\nabla}_2}\\
    \end{aligned}
    \ell_{0}(\nabla_i) =
        \begin{cases}
            \sign(\nabla_i) & \textrm{if } i=\argmax(\abs{\nabla})\\
            0 & \textrm{otherwise}
        \end{cases}
\end{equation*}

\noindent where \(\nabla\) is some gradient-like information. While any \lp{}-norm could be used in this layer, we also see natural extensions to other measurements of distance, such as LPIPS~\cite{zhang_unreasonable_2018} that could also fit into this component. This layer could also extend to allow for adaptive threat models, such what is used in the DDN attack~\cite{rony_decoupling_2019}.

\subsection{Travelers}

Travelers serve as the ``vehicles'' that navigate over a surface to meet
adversarial goals. Travelers are built from a series of subroutines that modify
\(x\): (1) \rr{}, (2) \cov{}, and (3) an optimization algorithm. Here, we
detail these components and describe how they aid in finding effective
adversarial examples.

\shortsection{Random-Restart} Many optimization problems, such as
\texttt{k-means}~\cite{hamerly_alternatives_2002} and
hill-climbing~\cite{russell_artificial_2009}, have been shown to benefit from
the meta-heuristic, \rr{}. Due to non-linear activation functions, deep neural
networks are non-convex, and thus, subject optimization algorithms to non-ideal
phenomena. Specifically, \rr{} attempts to prevent optimization algorithms from
becoming stranded in local minima by applying a random perturbation to an
input. At this time, \pgd{} is unique in its use of \rr{}, defined as:
\vspace{-1.5mm}
\begin{equation*}
    x = x + \mathcal{U}(-\epsilon, \epsilon)
\end{equation*}

\vspace{-1.5mm}
\noindent where \(\mathcal{U}\) is a uniform distribution, bounded by a
hyperparameter \(\epsilon\) (which represents the total perturbation budget).
Notably, while \rr{} could be applied at each perturbation iteration, \pgd{}
uses it once on initialization.

\shortsection{Change of Variables} As a new way of enforcing box
constraints,~\cite{carlini_towards_2017} introduced \cov{} for the \cwfull{}.
As noted in~\cite{carlini_towards_2017}, common practice for images is to first
scale features to be within \(\left[0,1\right]\). When a perturbation is
applied, these constraints must be enforced, as any feature beyond \num{1.0},
for example, would map to a pixel value greater than \num{255}, which exceeds
the valid pixel range for \qty{8}{\bit} images. Most attacks enforce this
constraint by simply clipping perturbations. However, this can negatively
affect certain gradient descent approaches~\cite{carlini_towards_2017}. Thus,
\cov{} was proposed to alleviate deficient behaviors. In the context of \cw{},
a variable \(w\) is defined and solved for (instead of the perturbation
directly). Its relationship to \(x\) is:
\vspace{-1.5mm}
\begin{equation*}
    x + \delta = \frac{1}{2}(\tanh{}(w) + 1)
\end{equation*}

\vspace{-1.5mm}
\noindent where \(\delta\) is the resultant perturbation applied to an input
\(x\). As~\cite{carlini_towards_2017} notes, this ensures that \(0\leq x +
\delta\leq 1\), meaning that examples will automatically fall within the valid
input range.

\shortsection{Optimizers} Nearly all attacks are described as
``taking steps in the direction'' (of a cost function). Practically
speaking, these attacks refer to \sgdfull{} (\sgd{}). As demonstrated by the
\bim{}, as little as three iterations (with \(\alpha=0.01\)) could be
sufficient to drop state-of-the-art \texttt{ImageNet} models to \(\qty{\sim
2}{\percent}\) accuracy~\cite{kurakin_adversarial_2016}. However, \cwfull{} was
perhaps the first attack to explicitly use \adam{} to craft adversarial
examples. \adam{}, unlike \sgd{}, adapts learning rates for every
parameter, and thus, often finds adversarial examples quicker than
\sgd{}~\cite{carlini_towards_2017}.

In addition to \sgd{} and \adam{}, we explore two additional optimizers, both
of which come from \autoattack{}. The first is \mbsfull{} (\mbs{}), which
accounts for momentum in its update step as follows:
\vspace{-1.5mm}
\begin{equation*}
    x_{i+1} = x_i + \eta\cdot\alpha\cdot\delta_i + (1-\eta)\cdot(x_i-x_{i-1})
\end{equation*}

\vspace{-1.5mm}
\noindent where \(\eta\) controls the strength of the momentum (set to \(0.75\)
in~\cite{croce_reliable_2020}). In addition to this momentum step, it also
features an adaptive learning rate that updates based on conditions that
capture progression of inputs toward adversarial goals, described
in~\cite{croce_reliable_2020}.

Finally, our framework also supports \bwsgdfull{} (\bwsgd{}), which is the
optimizer used for \fab{} in~\cite{croce_minimally_2020}. This optimizer
operates similarly to \sgd{} and \mbs{} but aims to update with the distance to
the original sample in mind by updating as follows:
\vspace{-1.5mm}
\begin{equation*}
    x_{i+1} = x_i + (1-\eta)\cdot\alpha\cdot\delta_i + \eta\cdot(x_{org} +
        \alpha\cdot \delta_{org})
\end{equation*}

\vspace{-1.5mm}
\noindent where \(\eta\) controls the influence of the original point on the
update step. In addition, if \(x_i\) is misclassified, this optimizer also
performs a backward step by moving \(x_i\) closer to \(x_{org}\) via: \(x_{i+1}
= \beta\cdot x_{i+1} + (1-\beta)\cdot x_{org}\). In our experiments, we set
\(\eta\) to be \num{0} since \(\delta_{org}\) (a) does not translate to attacks
that do not use a decision hyperplane projection and (b) as can be seen
in~\cite{croce_minimally_2020}, the use of backward step had a far greater
influence on the attack performance than setting a non-zero value of \(\eta\).

\section{Extending Performance Measurements}\label{methodology}

With the attack space made enumerable by our decomposition, we now focus on
necessary extensions of budget interpretations, the introduction of \oeafull{},
and our approach for measuring optimality.

\subsection{Beyond the \lp{}-norm}\label{methodology:beyond}

Since the inception of modern adversarial machine learning, the \textit{cost}
of producing an adversarial example has predominantly been measured through
\lp{}-norms. Yet, it seems impractical to assume realistic adversaries will be
unbounded by compute (as attacks that require days to produce adversarial
examples offer little utility in any real-time environment). This observation
is further exacerbated when attacks use expensive line search
strategies~\cite{szegedy_intriguing_2013}, embed hyperparameter optimization as
part of the crafting process~\cite{carlini_towards_2017}, or rely on model
Jacobian information~\cite{papernot_limitations_2016,
moosavi-dezfooli_deepfool_2016}. While such constructions can lead to
incredibly effective attacks, adversaries who are limited by compute resources
may find such attacks outright cost-prohibitive. To this end, we are motivated
to extend standard definitions of budget beyond exclusive measurements of
\lp{}-norms. Specifically, we incorporate and measure the \textit{time} it
takes to produce adversarial examples, therein extending our definition of
budget as:
\begin{equation}
    B(p, \theta, x) = \ell_{p}(x) + \theta\cdot T(x)
    \label{eq:budget}
\end{equation}

\noindent where \(p\) is the desired norm, \(\theta\) parameterizes the
importance of computational cost versus the introduced distortion, \(x\) is the
adversarial example, and \(T\) returns the compute time necessary to produce
\(x\). We note that the precise value of \(\theta\) depends on the threat
model; adversaries who are compute-constrained may prioritize time twice as
much as distortion (\ie{} \(\theta=2\)), while adversaries with strong compute
may not consider time at all (\ie{} \(\theta=0\), as is done in standard
evaluations). In \autoref{evaluation}, we find that some attacks consume
prohibitively large amounts of budget when compute is measured, and thus,
current threat models (which only measure \lp{} distance) fail to generalize
adversarial capabilities.

\subsection{Pareto Ensemble Attack}

With a realistic interpretation of budgets, we revisit a fundamental question:
\textit{Does an optimal attack exist?} Attacks measure distortion through
different \lp{}-norms, can require different amounts of compute, and have
varying budgets (which is notably true for robustness evaluations). Thus,
answering this question is non-trivial, especially in the absence of any
meaningfully large attack space.

A single definition that accurately characterizes optimality across attacks,
while incorporating these confounding factors, is challenging. Yet, we can say
some attack \(A\) is \textit{optimal} if, for a given threat model, \(A\)
bounds all other attacks for an adversarial goal (\ie{} \(A\) must lower-bound
all attacks when minimizing model accuracy across budgets). Of the \num{576}
attacks that we evaluated, no single attack met this definition. Thus, we
conclude that the optimal attack are best characterized by an \textit{ensemble
of attacks}.

\begin{figure}[t]
    \centering
    \resizebox{.85\columnwidth}{!}{\input{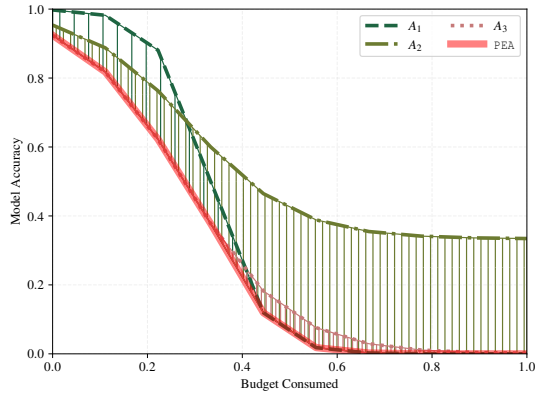}}
    \caption{The Optimal Attack --- The \oea{} lower-bounds all attacks across
    the range of budgets. Attacks \(A_1\) and \(A_3\) define the \oea{} for
    different budget ranges, while \(A_2\) is never part of the \oea{}. The
    area between the \oea{} and attack curves are shown with vertical
    bars.}\label{fig:oea}
\end{figure}

To this end, we introduce the \oeafull{} (\oea{}), a theoretical attack which,
for a given budget and adversarial goal, returns the set of adversarial
examples that best meet the adversarial goal, within the specified budget (in
other words, the Pareto frontier). The \oea{} is attractive for our analysis,
in that it serves as a meaningful baseline from which we can compare attack
performance to (discussed in the following section). Moreover, as an ensemble,
the \oea{} naturally evolves as the evaluated attack space expands. We formally
define the \oea{} as:
\begin{equation*}
    \oea{} = \bigcup_{b\in\mathscr{B}}\set{%
        \argmin_{x_{A\in\mathscr{A}}} \texttt{Acc}(f(x_A),\hat{y})\mid B(p,
        \theta, x_A)\leq b}
\end{equation*}

\noindent where \(b\) is a budget in a list of budgets \(\mathscr{B}\), \(x_A\)
is the set of adversarial examples produced by attack \(A\) from a space of
attacks \(\mathscr{A}\), \(f\) is a model, \(\hat{y}\) is the set of true
labels for \(x_A\), \(B\) is a function used to measure budget (\ie{}
\autoref{eq:budget}), \texttt{Acc} returns model accuracy, \(p\) is an
\lp{}-norm, and \(\theta\) controls the sensitivity to computational resources.
Concisely, the \oea{} returns the set of adversarial examples whose model
accuracy is minimal and within budget. Moreover, we provide a visualization of
the \oea{} in \autoref{fig:oea}, where the \oea{} forms the \textit{lower
envelope} of model accuracy across budgets.
We highlight that if there was some attack \(A'\) which achieved the lowest
accuracy across all budgets (for some domain), then the \(\oea{}=A'\). \iflinks\hypertargetblue{mr6a}{\mr{6a}}\fi It has
been suggested by some in the community that algorithms such as \pgd{} might be
optimal for some application~\cite{madry_towards_2017,
zheng_distributionally_2019, carlini_provably_2017}. Our formulation of the
\oea{} and measure of optimality allows us to test this hypothesis.

\shortsection{Measuring Optimality}\label{methodology:measuring} The \oea{} yields a baseline from which we can fairly assess the performance of
attacks. As the \oea{} meets the definition of \textit{optimal} (that is, it
bounds attack performance), we can evaluate attack performance relative to the
\oea{}. Intuitively, attacks that closely track the \oea{} are performant,
while those that do not are suboptimal. Mathematically, this can be measured as
the area between the curves of the \oea{} and some attack \(A\). \iflinks\hypertargetblue{mr4c}{\mr{4c}}\fi We note that
our definition of optimality is: (1) relative to the attacks considered (and
not measured against a set of provably worst-case adversarial examples or
certified robustness~\cite{richardson_bayes-optimal_2021, vos_robust_2021,
carlini_provably_2017}), and (2) as attacks are ranked by area, prefers attacks
that are consistently performant (\ie{} across the budget space). We
acknowledge this measurement favors attacks whose behaviors are stable (which
we argue most popular white-box attacks exhibit); other modalities may benefit
from other cost measures.

For example, in \autoref{fig:oea}, the area between the \oea{} and attack
\(A_2\) is maximal, minimal for attack \(A_3\), and somewhere in between for
attack \(A_1\). Thus, we conclude that the worst-case adversary would use
\(A_3\) if bound by small budgets, otherwise \(A_1\) (and never \(A_2\)). This
approach to measuring attack performance is desirable in that, (1) attacks that
track the \oea{} across budgets have minimal area (and thus, constitute a
performant attack), and (2) attacks that are exclusively optimal for specific
budgets incur large area, which allows us to differentiate attacks that are
\textit{always} performant from those that are \textit{sometimes} performant.

\section{Evaluation}\label{evaluation}

With our attack decomposition and
approach to measure optimality, we ask several questions: (1) Do known attacks
perform best? (2) What attacks are optimal, if any? (3) Which components tend
to yield performant attacks?

\subsection{Setup}\label{evaluation:setup}

We perform our experiments on a \texttt{Tensor EX-TS2} with two \texttt{EPYC
7402} CPUs, \qty{1}{\tebi\byte} of memory, and four \texttt{Nvidia A100} GPUs.
We use \pytorch{}~\cite{paszke_pytorch_2019} \texttt{1.9.1} for instantiating
learning models and our attack decomposition.
Here, we describe the attacks, threat models, robustness approach (\ie{}
adversarial training), and datasets used in our evaluation. We defer attack
adaptations to \aref{appendix-e:modifications}, and hyperparameters \& details
on adversarial training to \aref{appendix-e}. \ifarxiv\else Full versions of any shortened tables and figures in this section can be found in the arxiv version of this paper~\cite{}.\fi 

\shortsection{Attacks} In \autoref{decomposition}, we introduce a decomposition
of adversarial machine learning by atomizing attacks into modular components.
\iflinks\hypertargetblue{mr4e}{\mr{4e}}\fi Our evaluation spans the enumerated
\num{576} attacks. Of these \num{576}, \jsma{}, \cw{}, \df{}, \pgd{}, \bim{},
\apgdce{}, \apgddlr{}, and \fab{} are labeled explicitly, while other attacks
are numbered from \num{0} to \num{575}. The specific component choices of
attacks mentioned by number can be found in \ifarxiv\aref{appendix-b:encoding}\else\aref{appendix-e}\fi. We note
that some known attacks (such as \df{} and \cw{}) have specialized variants for
\lp[p\neq2]-norms, which we do not implement (as to maintain homogeneous
behaviors across attacks of the same norm). Thus, we still reference these
attacks numerically, since they are not the algorithmically identical.

In our experiments, we focused on untargeted attacks: that is, the adversarial
goal is to minimize accuracy. While our decomposition is readily amenable to
targeted variants, we defer analysis (and thus evaluation) of targeted attacks
for two reasons: (1) choosing a target class requires domain-specific
justification, and (2) certain classes are harder to attack than
others~\cite{papernot_limitations_2016}. These two factors would require a
rather nuanced analysis, while our objectives aim to characterize broad attack
behaviors. Thus, we anticipate that while a targeted analysis might affect
attack performance in an absolute sense, relative performance to other attacks
will likely be indifferent.

\shortsection{Threat Models} As motivated in \autoref{methodology}, we explore
the interplay in attack performance when compute is measured, as defined by
\autoref{eq:budget}. Specifically, we explore \num{3} \lp{}-based threat models
(\ie{} \lp[0], \lp[2], and \lp[\infty]) with \num{20} different values of
\(\theta\) at \num{0.1} step sizes, from \numrange{0}{2}. These values can be
interpreted as an adversary who, for example, values computational speed
\textit{twice as much} over minimizing distortion (\ie{} \(\theta=2\)). We
note that all attacks are instantiated within our framework, and thus, any
implementation-specific optimizations that accelerate compute speed 
are leveraged uniformly across attacks.

\shortsection{Robust Models} \iflinks\hypertargetblue{mr4d}{\mr{4d}}\fi
Adversarial training~\cite{madry_towards_2017, goodfellow_explaining_2014} is
one of the most effective defenses against adversarial examples to
date~\cite{croce_robustbench_2020, shafahi_adversarial_2019,
carlini_provably_2017}. Given its popularity and compelling results, we are
motivated to investigate the impact of robust models on relative attack
performance. We adversarially train our models with a \pgd{}-based adversary.
We follow the same approach as shown in~\cite{madry_towards_2017}: input
batches are replaced by adversarial examples (produced by \pgd{}) during
training. For \mnist{} and \cifar{}, hyperparameters were used 
from~\cite{madry_towards_2017}; other datasets were trained with parameters
which maximized the accuracy over benign inputs and adversarial examples.
Additional hyperparameters can be found in \aref{appendix-e}.

\subsubsection{Datasets} \iflinks\hypertargetblue{mr2}{\mr{2}}\fi We use seven
different datasets in our experiments, chosen for their variation across
dimensionality, sample size, and phenomena. We provide details and basic
statistics below.

\shortsection{\phishing{}} The \phishing{}~\cite{chiew_new_2019} dataset is
designed for detecting phishing websites. Features were extracted from
\num{5000} phishing websites and \num{5000} legitimate webpages. It contains
\num{48} features and \num{10000} samples. Beyond its phenomenon, we use the
\phishing{} dataset to investigate the effects of small dimensionality and
training size on attack performance.

\shortsection{\nslkdd{}} The \nslkdd{}~\cite{tavallaee_detailed_2009} is based
on the seminal \texttt{KDD Cup '99} network intrusion detection dataset.
Features are defined from varying network features from traffic flows emulated
in a realistic military network. At \num{41} features, it contains \num{125973}
samples for training and \num{22544} for testing. We use the \nslkdd{} for its
small dimensionality, large training size, and concept
drift~\cite{gomes_adaptive_2017}.

\shortsection{\unswnb{}} The \unswnb{}~\cite{moustafa_unsw-nb15_2015} is a
network intrusion detection dataset designed to replace the \nslkdd{}. Features
are derived from statistical and packet analysis of real innocuous flows and
synthetic attacks. It has \num{48} features, with \num{175341} samples for
training and \num{83332} samples for testing. The \unswnb{} enables us to
compare if attacks generalize to similar phenomenon (such as the \nslkdd{}).

\shortsection{\malmem{}} \malmemfull{}
(\malmem{})~\cite{carrier_detecting_2022} is a modern malware detection
dataset. \num{58} features are extracted from memory dumps of benign
applications and three different malware families (\ie{} trojans, spyware, and
ransomware). In total, it contains \num{58,596} samples, with half belonging to
benign applications and half to malware. \malmem{} gives us the opportunity to
understand the effects of small dimensionality in an entirely different
phenomenon from the network datasets.

\shortsection{\mnist{}} \mnist{}~\cite{lecun_gradient-based_1998} is a dataset
for handwritten digit recognition. It is a well-established benchmark in
adversarial machine learning applications. With \num{784} features, \num{60000}
samples for training and \num{10000} for testing, \mnist{} has substantially
larger dimensionality than even the largest network datasets. We use \mnist{}
to corroborate prior results, explore a vastly different phenomenon, and
investigate how (relatively) large dimensionality influences attack
performance.

\shortsection{\fmnist{}} \fmnistfull{}
(\fmnist{})~\cite{xiao_fashion-mnist_2017} is a dataset for recognizing
articles of clothing from Zalando articles. Advertised as a drop-in replacement
for \mnist{}, \fmnist{} was designed to be a harder task and closer
representative of modern computer vision challenges. \fmnist{} has identical
dimensionality, training samples, and test samples to \mnist{}. Thus, we use
\fmnist{} to understand if changes in phenomena alone are sufficient to
influence attack performance.

\shortsection{\cifar{}} \cifar{}~\cite{krizhevsky_learning_2009} is a dataset
for object recognition. Like \mnist{}, \cifar{} is extensively used in
adversarial machine learning literature. At \num{3072} features, \cifar{}
represents a substantial increase in dimensionality from \mnist{}. It has
\num{60000} samples for training and \num{10000} for testing. \cifar{} allows
us to compare against extant works and explore how domains with extremely large
dimensionality affect attack efficacy.

\subsection{Comparison to Known Attacks}\label{eval:known}

As discussed in \autoref{decomposition}, we contribute \num{568} new attacks.
Naturally following, we ask: \textit{are any of these attacks useful?} Asked
alternatively, \textit{do known attacks perform best?} We investigate this
question through commonly accepted performance
measurements~\cite{carlini_adversarial_2017, papernot_limitations_2016,
moosavi-dezfooli_deepfool_2016, kurakin_adversarial_2016}: the amount of \lp{}
budget consumed by attacks whose resultant adversarial examples cause model
accuracy to be \qty{<1}{\percent}. In this traditional performance setting, we
aim to understand if known attacks serve as the Pareto frontier (which would
indicate that our contributed attacks yield little in terms of adversarial
capabilities).

We organize our analysis as follows: (1) we first segment attacks based on
\lp{}-norm and compare them to known attacks of the same norm (that is, we
compare \jsma{} to \lp[0] attacks, \cw{}, \df{}, \& \fab{} to \lp[2], and
\pgd{}, \bim{}, \apgdce{}, \& \apgddlr{} to \lp[\infty]), and (2) report
relative budget consumed (with respect to known attacks) for attacks whose
adversarial examples caused model accuracy to be \qty{<1}{\percent}.

\subsubsection{Performance on \mnist{}} 
\begin{figure*}[t]
    \centering
    \begin{minipage}{0.25\textwidth}
        \resizebox{\textwidth}{!}{\includegraphics{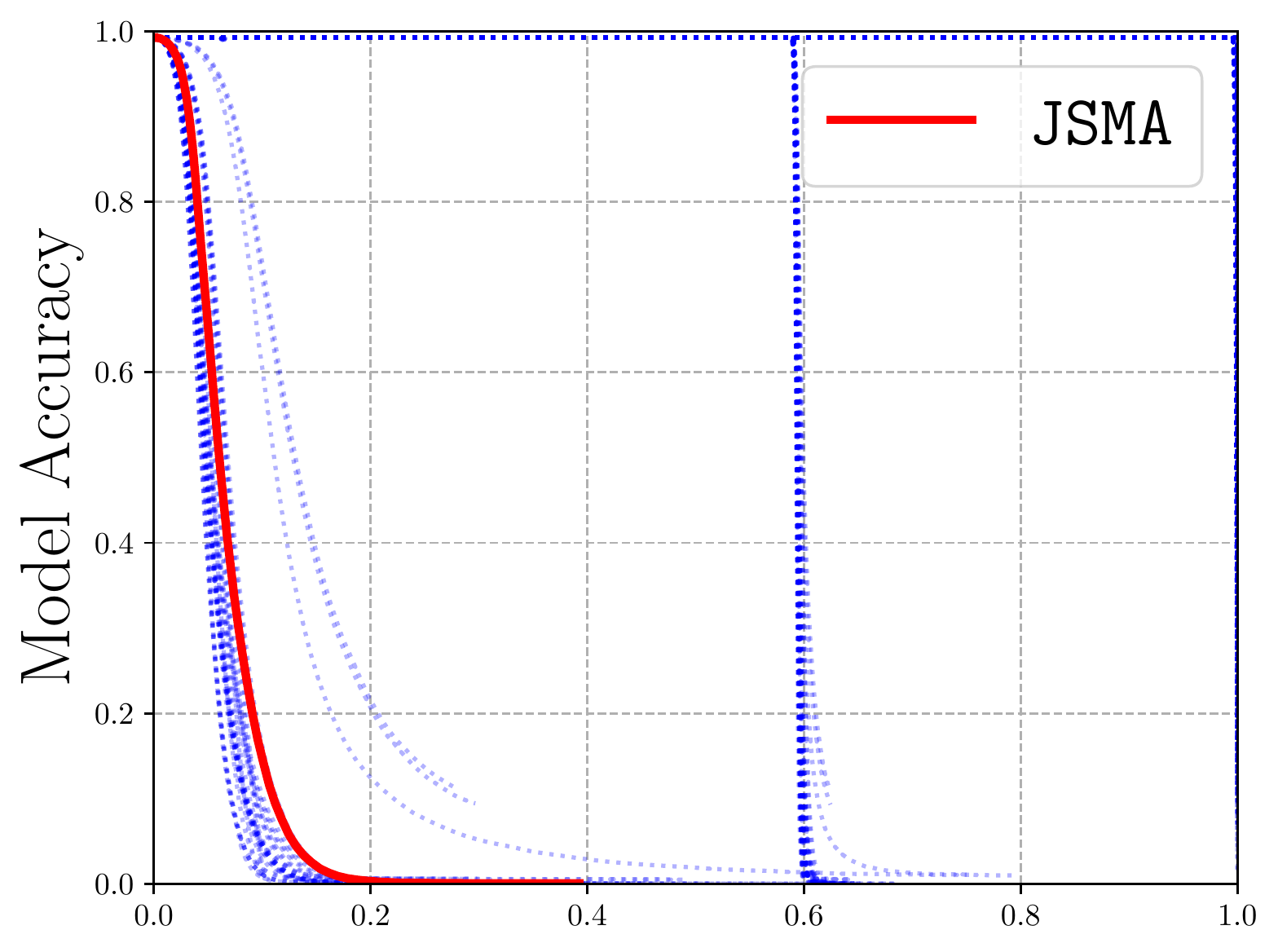}}
        \subcaption{\(\lp[0]\)}\label{fig:mnist_l00time}
    \end{minipage}
    \begin{minipage}{0.25\textwidth}
        \resizebox{\textwidth}{!}{\includegraphics{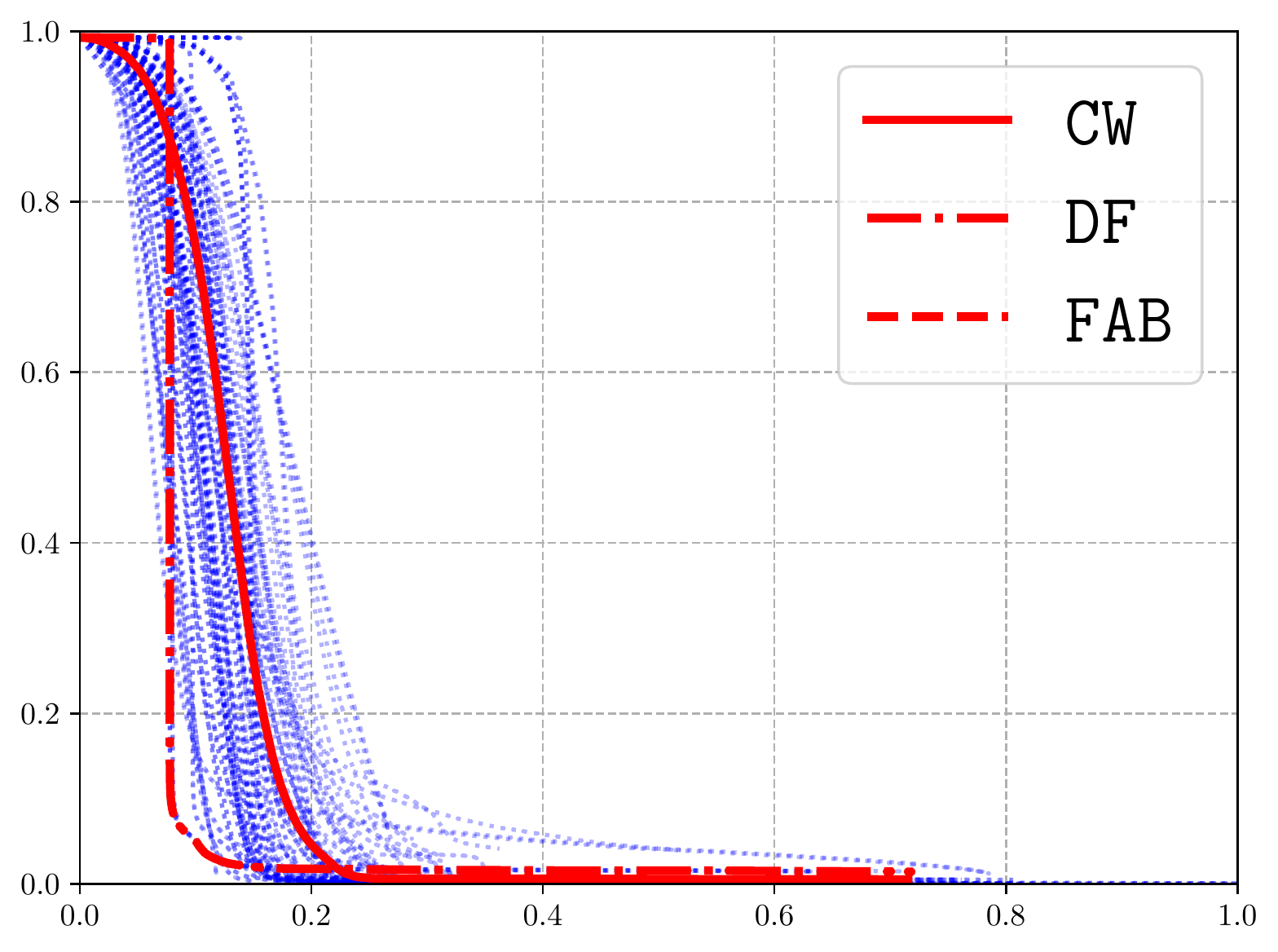}}
        \subcaption{\(\lp[2]\)}\label{fig:mnist_l20time}
    \end{minipage}
    \begin{minipage}{0.25\textwidth}
        \resizebox{\textwidth}{!}{\includegraphics{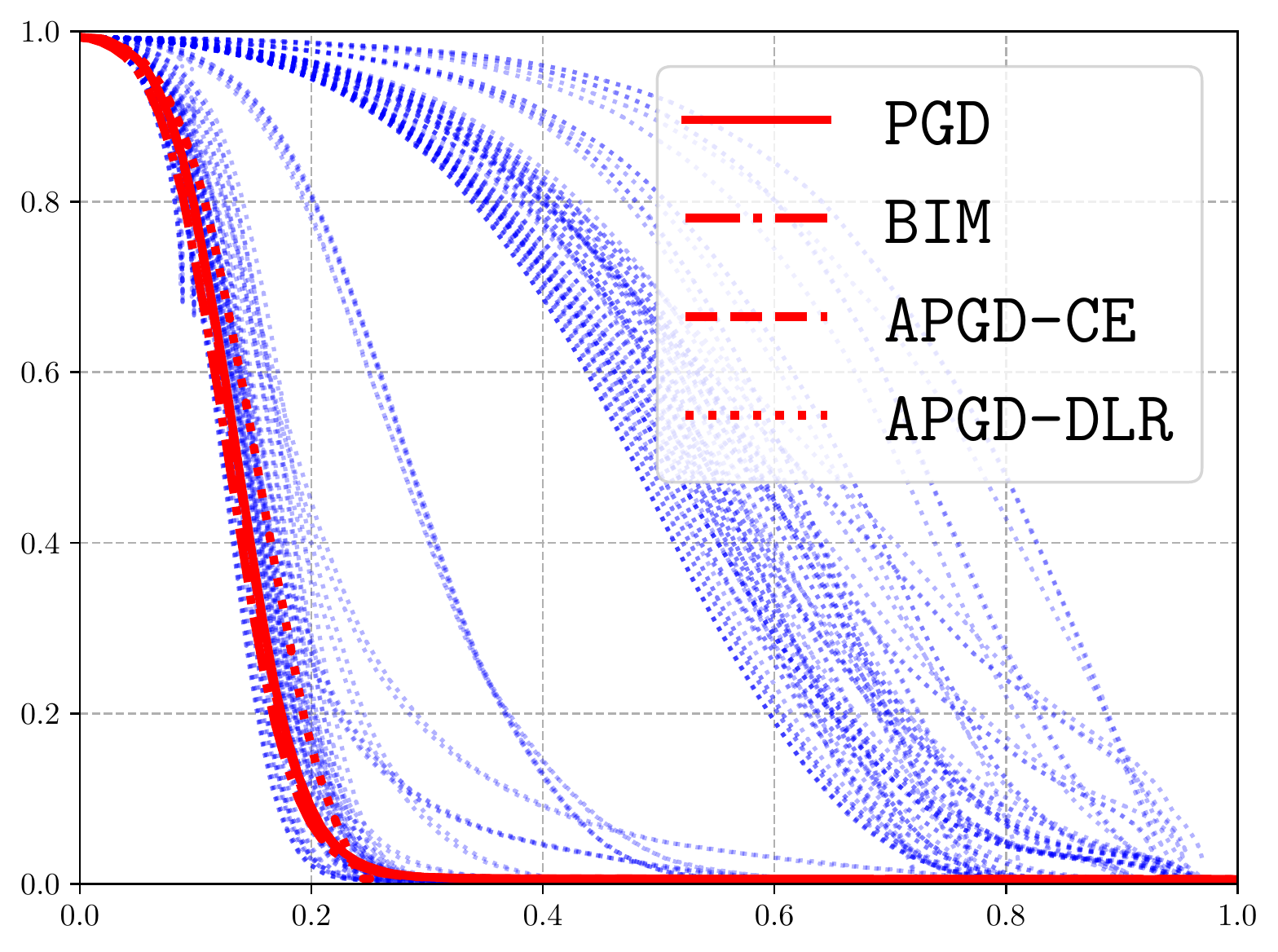}}
        \subcaption{\(\lp[\infty]\)}\label{fig:mnist_linf0time}
    \end{minipage}
    \caption{\mnist{} model accuracy for (normalized) \lp[0]-, \lp[2]-, and
    \lp[\infty]-based budgets. Known attacks are highlighted in red. Results
    show median accuracy across \num{1000} iterations over \num{10}
    trials.}\label{fig:mnist_comparison}
\end{figure*}

For our analysis of attack performance,
we craft adversarial examples for \num{1000} iterations over ten trials (we
note that \num{1000} iterations was selected for completeness; the vast
majority of attacks converged in less than \num{100} iterations).
\autoref{fig:mnist_comparison} shows the median results for two threat models,
segmented by \lp{}-norm. Known attacks (\ie{} \jsma{}, \cw{}, \df{} \fab{},
\pgd{}, \bim{}, \apgdce{}, and \apgddlr{}) are highlighted in red, while other
attack curves are dotted blue and slightly opaque to capture density. We now
discuss our results on a per-norm basis.

\shortsection{\lp[0] Attacks} \autoref{fig:mnist_l00time} shows \lp[0]-targeted
attack performance with the \jsma{} in red. We observe that the \jsma{}
is worse than most attacks. Attack performance is largely well-clustered with a
few poor performing attacks near the top right portions of the graph. These
attacks used \rr{}, and thus, immediately consume most of the available \lp[0]
budget.

\shortsection{\lp[2] Attacks} \autoref{fig:mnist_l20time} shows \lp[2]-targeted
attacks, with \cw{} as solid red, \df{} as dash-dotted red, and \fab{} as
dashed red. Like \lp[0], attacks are well-concentrated (albeit with slightly
more spread). Notably, \df{} and \fab{} (which are ostensibly superimposed on
one another), demonstrate impressive performance (the red lines that are nearly
vertical)---both drop model accuracy with a near-zero increase in budget. \cw{}
exhibits moderate performance over the budget space.

\shortsection{\lp[\infty] Attacks} \autoref{fig:mnist_linf0time} shows
\lp[\infty]-targeted attacks, with \pgd{} as solid red, \bim{} as dash-dotted
red, \apgdce{} as dashed red, and \apgddlr{} as dotted red. Unlike other norms,
\lp[\infty] has clear separation, broadly attributable to using \cov{}
(specifically, attacks that used \cov{} performed worse than those that did
not). Finally, all of the known attacks exhibit near-identical performance,
with \apgddlr{} slightly pulling ahead at budgets \(>0.2\).

From our norm-based analysis, we highlight that: (1) \rr{} is largely
inappropriate for \lp[0]-targeted attacks (in that benefits do not outweigh the
cost), (2) \lp[2]-targeted attacks cluster fairly well; no individual attack
substantially outperformed any other, and (3) \lp[\infty]-targeted attacks were
broadly unable to exploit \cov{}.

\subsubsection{Relative Performance to Known Attacks} Recall our central
question for this experiment: \textit{do known attacks perform best?} To answer
this question, we analyze the minimum budget necessary for attacks to cause
model accuracy to be \qty{<1}{\percent} (attacks that fail to do so are encoded
as consuming infinite budget). We run attacks for \num{1000} iterations over
ten trials and report the median results in \autoref{tab:mnist_comparison}.
\begin{table*}[!t]
    \centering
    \resizebox{.9\textwidth}{!}{%
    \begin{tabular}{rccc rccc rccc}
        \toprule
        \multicolumn{4}{c}{\textbf{\lp[0] Attacks}} &
        \multicolumn{4}{c}{\textbf{\lp[2] Attacks}} &
        \multicolumn{4}{c}{\textbf{\lp[\infty] Attacks}} \\
        \cmidrule(lr){1-4}\cmidrule(lr){5-8}\cmidrule(lr){9-12}
        Rank & Attack & \% Reduction & \lp[0] Budget      &
        Rank & Attack & \% Reduction & \lp[2] Budget      &
        Rank & Attack & \% Reduction & \lp[\infty] Budget \\
        \midrule
        1.   & \fw[171] & -\qty{41}{\percent} & 0.10 &
        1.   & \fw[460] & -\qty{50}{\percent} & 0.12 &
        1.   & \fw[449] & -\qty{8}{\percent}  & 0.22 \\

        30.  & \textbf{\jsma{}}    & --- & 0.17 &
        69.  & \textbf{\cw{}}      & --- & 0.24 &
        17.  & \textbf{\apgddlr{}} & --- & 0.24 \\

        68.  & \fw[246]        & +\qty{352}{\percent} & 0.77 &
        88.  & \fw[37]         & +\qty{45}{\percent}  & 0.35 &
        40.  & \textbf{\bim{}} & +\qty{16}{\percent}  & 0.28 \\

        \multicolumn{4}{c}{}      &
        136. & \textbf{\df{}}     & \(+\infty\%\)       & \(\infty\) &
        48.  & \textbf{\apgdce{}} & +\qty{20}{\percent} & 0.29 \\

        \multicolumn{4}{c}{}   &
        137. & \textbf{\fab{}} & \(+\infty\%\)       & \(\infty\) &
        49.  & \textbf{\pgd{}} & +\qty{20}{\percent} & 0.29 \\

        \multicolumn{4}{c}{}      &
        \multicolumn{4}{c}{}      &
        135. & \fw[191] & +\qty{304}{\percent} & 0.97 \\
        \bottomrule
    \end{tabular}}
    \caption{\mnist{} Relative Attack Comparisons. Budgets are normalized.
    Attacks that fail to reduce model accuracy to be \qty{<1}{\percent}
    are labeled as consuming infinite budget. Budget reductions are relative
    to the best known attack for each \lp{}-norm.}\label{tab:mnist_comparison}
\end{table*}
Here, attacks are ranked by budget and segmented
by norm (\ie{} \num{192} attacks per norm). We report the percentage change of
each attack with respect to the known attack that performed best in that norm
(that is, for \lp[0], results are relative to the \jsma{}, while for \lp[2],
results are relative to \cw{}, which outperformed \df{}, etc.). In the table, we
show: (1) the attack that ranked first, (2) ranks of known attacks, and (3) the
lowest ranked attack that still reduced model accuracy to \qty{<1}{\percent}.
Next, we highlight some strong trends for each \lp{}-norm. 

Of the \qty{34}{\percent} of attacks that succeed in the \lp[0] space, the
\jsma{} (ranked \(32^\mathrm{nd}\)) was at the bottom of the highly performant
bin (in that its \lp[0] budget was \num{0.17})---the \jsma{} was held back by
its saliency map, \jsmasmap{}; using either \dfsmap{} (or no saliency map at
all, \ie{} \identitysmap{}) was almost always better. While \cw{} seemingly
rank low (\ie{} \(69^\mathrm{th}\)), we note that \lp[2] budgets were broadly
similar, as the worst and best performing attacks were within
\(\pm\)\qty{50}{\percent} of the budget consumed by \cw{}. \apgddlr{}, \bim{},
\apgdce{}, and \pgd{}, ranked \(17^\mathrm{th}\), \(40^\mathrm{th}\),
\(48^\mathrm{th}\), and \(49^\mathrm{th}\) respectively, were marginally
outperformed by attacks using either the \dfsmap{} saliency map or \bwsgd{}
optimizer. As a final note, we were confounded by the performance of \df{} and
\fab{}---visually inspecting \autoref{fig:mnist_l20time}, both are
\textit{clearly} superior attacks (the performance curves ostensibly resemble
square waves) and yet, they failed to reduce model accuracy to
\qty{<1}{\percent}. While these analyses of attack performance has been useful
historically for understanding adversarial examples, we argue that this ``race
to \qty{0}{\percent} accuracy'' fails to capture meaningful definitions of
attack performance (as made evident by the apparent ``failure'' of \df{} and
\fab{}).

From our comparison with known attacks, we highlight two key takeaways: (1)
Measuring the required distortion to reach some amount of model accuracy is a
rather crude approach to estimating attack performance. We argue using
measurements that factor the entire budget space (such as the \oea{}, which we
use subsequently) will yield more meaningful interpretations of attack
performance. (2) Even when we define success as \qty{<1}{\percent} model
accuracy, known attacks do not perform best. In fact, many attacks produced by
our decomposition consistently outperformed known attacks (\eg{} \num{68} out
of the \num{189} introduced by our approach outperformed both \cw{} and \df{}),
which demonstrates the novel adversarial capabilities introduced by our
decomposition. 

\begin{figure}[t!]
    \centering
    \resizebox{0.42\textwidth}{!}{\includegraphics{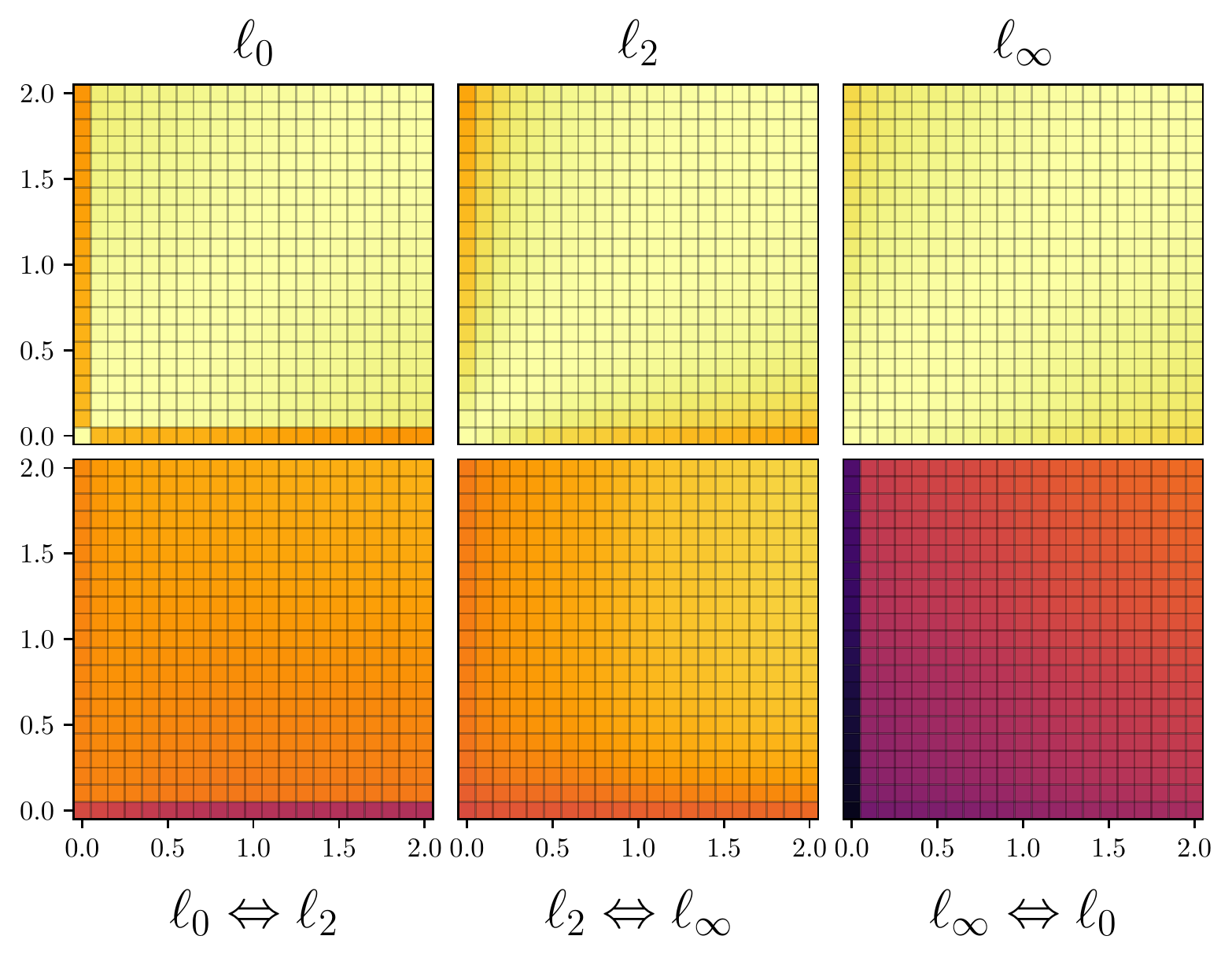}}
   \raisebox{0.2in}{%
   \resizebox{!}{2.05in}{\includegraphics{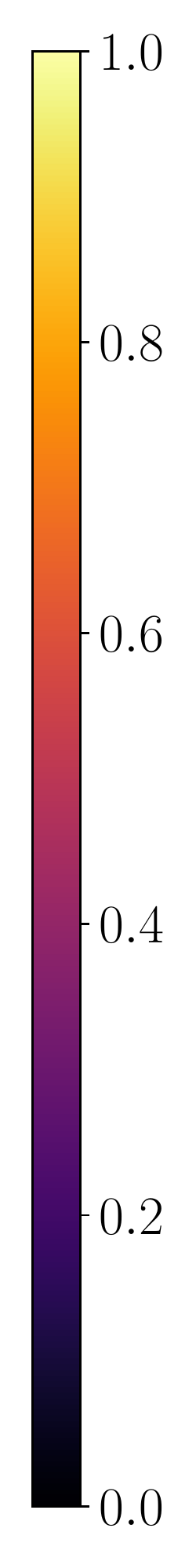}}}
    \caption{Median Spearman Rank Correlation Coefficients for
    \mnist{}---Results are segmented by \lp{}-norm. Data points correspond to a
    specific threat model (\ie{} a value for \(\theta\)). 
    High attack performance
    generalization is encoded as lighter shades, while low generalization is
    encoded with darker shades. Top row shows within-norm generalization \&
    bottom row shows cross-norm generalization.}\label{fig:spearman_mnist}
\end{figure}

\subsection{Optimal Attacks}\label{eval:optimal}

In \autoref{methodology:measuring}, we introduced an approach for measuring
optimality: the area between the performance curves of the \oea{} and an
attack. Attacks that have a small area closely track the \oea{} and thus, are
performant attacks, while those that have a large area perform poorly. In this
experiment, we ask: \textit{does attack performance generalize?} In other
words, is relative attack performance invariant to dataset or threat model?

We investigate this hypothesis of attack optimality by ranking attacks by area
across varying threat models, datasets, and robust models. Then, we measure the
generalization of these rankings via the Spearman rank correlation
coefficient~\cite{spearman_proof_1987},
which informs us how similar the rankings are between two datasets, threat
models, or a robust and non-robust model.

For example, a highly positive correlation across two datasets would imply that
relative attack performance was unchanged (in other words, changing the dataset
had little to no effect on attack performance), a near-zero correlation would
suggest that relative attack performance changed substantially (which would
suggest attack performance is sensitive to the dataset), and a negative
correlation would indicate attack performance was reversed (\ie{} the worst
attacks on one dataset became the best on another). In our experiments, we
craft adversarial examples for \num{1000} iterations over ten
trials\footnote{In another experiment, we validated that rankings are highly
correlated across trials. Combined with our use of nonparametric statistics
(\ie{} median Spearman correlation), this ensures our metrics converge in few
trials.} and report the median Spearman rank correlation coefficients. We note
that \num{1000} iterations was selected for completeness; the vast majority of
attacks converged in less than \num{100} iterations.

\shortsection{Optimal Attacks by Threat Model} Here, we analyze the
generalization of attack performance \textit{across threat models}.
Specifically, we consider \lp[0]-, \lp[2]-, and \lp[\infty]-based threat models
with varying values of \(\theta\) (from \numrange{0}{2}). Note that
\(\theta=0\) (\ie{} where compute time is ignored) is the commonly used threat
model. \autoref{fig:spearman_mnist} shows the median Spearman rank correlation
coefficients for \mnist{} \ifarxiv(other datasets are listed in \aref{appendix-d})\fi with
results segmented by \lp-norm{}. Each entry corresponds to a unique threat
model (\ie{} a value for \(\theta\)). High attack performance generalization is
encoded as lighter shades, while low generalization is encoded with darker
shades.

From the results, we can readily observe: (1) rankings do not generalize across
\lp{}-norms, especially between \lp[0]- and \lp[\infty]-targeted attacks (but
do generalize relatively well \textit{within} an \lp{}-norm), and (2) the
influence of compute on rankings appears to be \lp{}-norm dependent:
\lp[0]-based threat models that weight compute (\ie{} \(\theta\neq 0\)) do not
generalize well to those that do not, \lp[2]-based threat models exhibit a
smoother degradation of generalization, while, surprisingly, \lp[\infty]-based
threat models generalize \textit{everywhere} (that is, the same attacks that
were found to be performant with \(\theta=2\) were as performant when
\(\theta=0\)). Within a dataset, we observe that the threat model significantly
affects attack performance across \lp{}-norms, and to some extent, within an
\lp{}-norm, with \lp[\infty] as the exception.

\shortsection{Optimal Attacks by Dataset} In this experiment, we instead now
measure the generalization of attack performance \textit{across datasets}.
Specifically, for a given threat model, we measure the generalization of attack
performance rankings across seven datasets. The evaluated datasets span varying
forms of phenomena, from classifying network traffic to categorizing clothing
items, and thus, we investigate if performant attacks are task-agnostic.
\ifarxiv Correlations for threat models with \(0<\theta<2\) can be found in
\aref{appendix-d}.\fi

The results in \autoref{fig:spearman_tm} disclose that: (1) \cifar{} does not
generalize at all, regardless of \lp{}-norm, (2) skewing budgets towards
favoring compute gradually degrades generalization---attack
rankings become increasingly dissimilar as we move from ignoring compute time
(\(\theta=0\)) to heavily favoring it (\(\theta=2\)), and (3) \lp[0]-based
threat models readily generalize across datasets and is largely invariant to
considering compute, \lp[2] attacks, regardless of \(\theta\), moderately
generalize, and \lp[\infty] attacks closely track \lp[2] attacks with a
particular subtly: attacks performant on \mnist{} generalized almost perfectly
to \fmnist{} (\ie{} image-based generalization), while attacks performant on
\nslkdd{} almost perfectly generalized to \unswnb{} (\ie{}
network-intrusion-detection-based generalization). Lastly, we observe that
attacks performant on \phishing{} and \malmem{} moderately generalized
better to non-image data (particularly to the \unswnb{}). Considering compute
degrades these observations slightly. Within a threat model, we observe that,
based on \lp{}-norm, the dataset can have drastic degrees of influence on
attack performance, in that it can have little effect at all (\eg{} \lp[0]),
have an effect everywhere (\ie{} \lp[2]), or have an effect specific to the
phenomena (\ie{} \lp[\infty]). We attribute the unique behavior of \cifar{} to
its dimensionality; the next largest dataset, \fmnist{} and \mnist{}, are
\(\sim\)\qty{74}{\percent} smaller.

\shortsection{Optimal Attacks Against Robust Models} In this final experiment,
we now measure the generalization of attack performance \textit{between robust
and non-robust models}. Specifically, for a given threat model and dataset, we
compute pairwise correlations between attack performance rankings on robust and
non-robust models. Adversarially trained models have been shown to be an
effective defense against adversarial examples~\cite{madry_towards_2017}, and
thus, we investigate if such procedures have a visible effect on attack
performance.

\begin{figure}[t]
    \centering
    \resizebox{.42\textwidth}{!}{\includegraphics{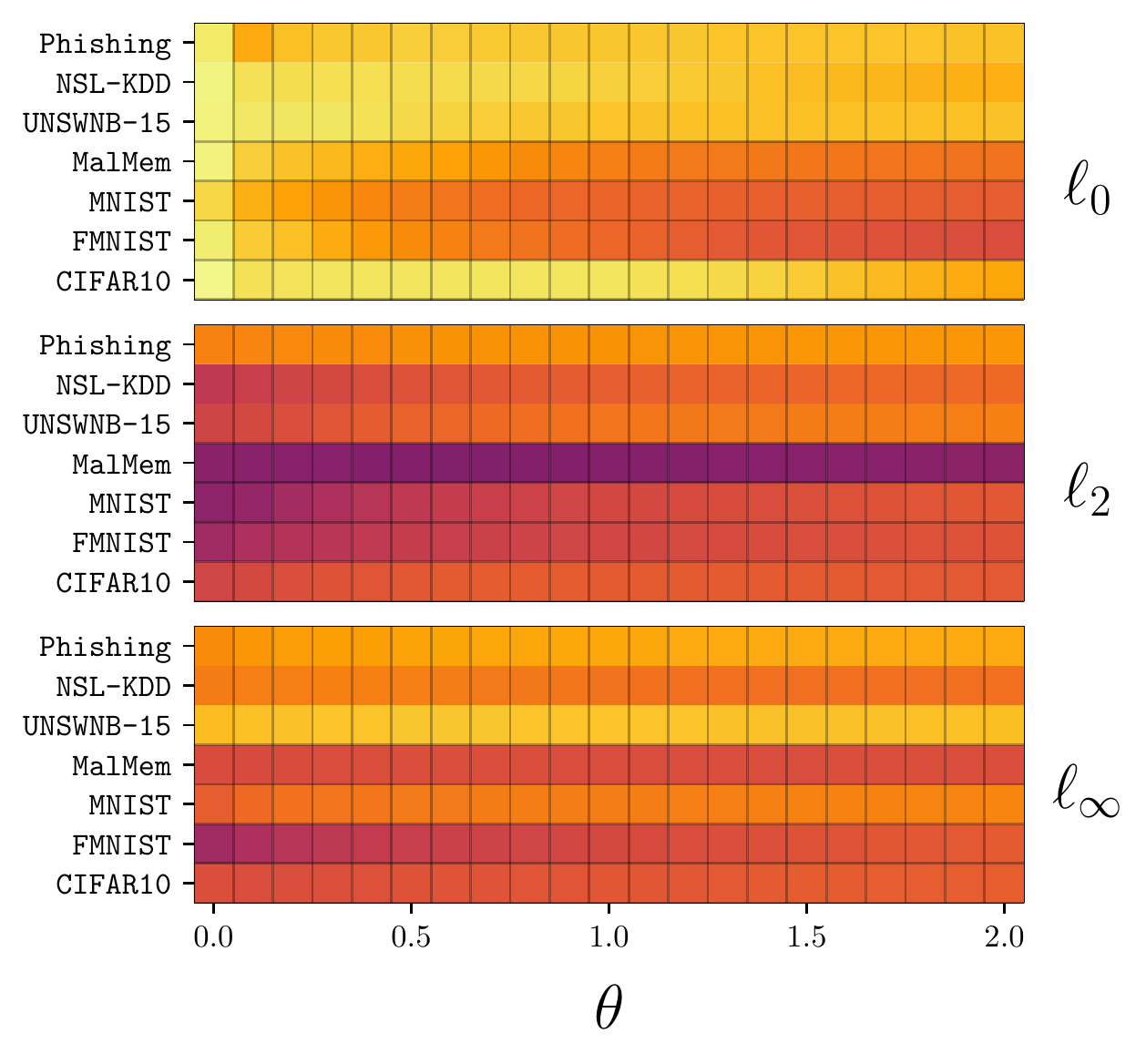}}
    \resizebox{!}{2.7in}{\includegraphics{figures/spearman_vcolorbar.pdf}}
    \caption{Median Spearman Rank Correlation Coefficients for robust and
    non-robust models---Results are segmented by \lp{}-norm. Data points
    correspond to a specific threat model (\ie{} a value for $\theta$). Lighter
    shades show high generalization of attack performance ranking across that
    dataset and threat model, while darker shades are low
    generalization.}\label{fig:spearman_robust}
\end{figure}

Median Spearman rank correlation coefficients for all datasets and threat
models are shown in \autoref{fig:spearman_robust}. We note several trends
across norm, threat models, and datasets: (1) generally speaking, attack
rankings in \lp[2]-based threat models were \textit{substantially} affected by
robust models, especially for~\malmem{},~\mnist{}, and~\fmnist{}, (2)
considering compute can have a significant impact on generalization, mainly
dependant on the norm; increasing the importance of compute almost universally
aided generalization in \lp[2], but hurt generalization in \lp[0] (especially
for image data, albeit \cifar{} is less sensitive to varying \(\theta\) at the
scales we investigated), and (3) we observed that top-performing attacks can be
especially affected: on \mnist{} for an \(\lp[2]+0\cdot\mathrm{time}\) threat
model, for instance, the top \num{10} attacks on the non-robust model had a
median rank of \(445^{th}\) (out of \num{576}) on the robust model. These
profound differences in relative attack effectiveness demonstrate that the
unique properties of robust models necessitate changes to attack components (discussed further in \autoref{explainability:robust}).

\shortsection{Takeaways on Attack Optimality} In this set of experiments, we
analyzed attack optimality through the lens of varying threat models, unique
data phenomena, and robust models. From our analyses, we find that the
optimality of any given attack is highly dependant on the given context. We
support this conclusion through the following remarks on the generalization of
relative attack performance: (1) across threat models, performance generalizes
well within an \lp{}-norm, but not across---considering compute exacerbates
this observation, (2) across datasets, performance generalization is broadly
sensitive to \lp{}-norm (with \cifar{} generalizing poorly everywhere), and (3)
between robust and non-robust models, attack rankings are largely a function of
data phenomena (\eg{} image-based phenomena exhibit poor generalization,
regardless of the threat model).

\begin{figure}[t]
    \centering
    \resizebox{.42\textwidth}{!}{\includegraphics{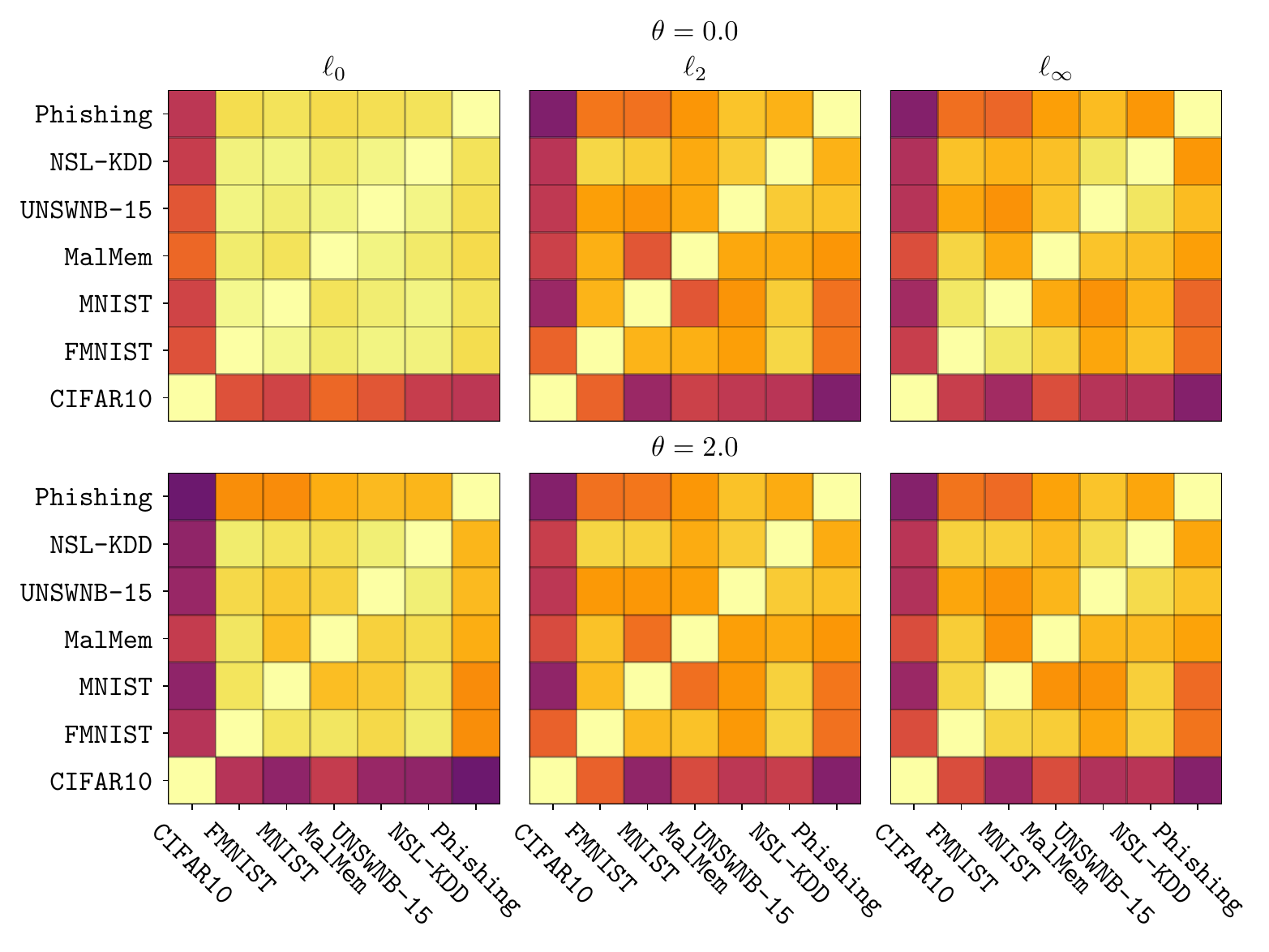}}
    \resizebox{!}{2.05in}{\includegraphics{figures/spearman_vcolorbar.pdf}}
    \caption{Median Spearman Rank Correlation Coefficients for \(\theta=0\) and
    \(\theta=2\) threat models---Results are segmented by \lp{}-norm. Entries
    correspond to a dataset pair. High attack performance generalization is
    encoded as lighter shades, while low generalization is encoded with darker
    shades.}\label{fig:spearman_tm}
\end{figure}

\subsection{When and Why Attacks Perform Well}

With our metric for attack performance established and evaluated, we proceed by
asking, \textit{why do certain attacks perform well?} Here, we explore the
general trends of attack components and their influence on performance through
a series of hypothesis tests. We build a space of possible hypotheses of
relative attack performance (over all attack components), perform hypothesis
testing against this space, and identify those with the highest significance
and effect size. We begin with significant hypotheses of non-robust models and
conclude with hypotheses most affected by model robustness.

\subsubsection{The Space of Hypotheses} We define a \textit{hypothesis} as a
comparison between two component values (which we label as \(H_1\) and
\(H_2\)), such as \textit{``using \crossentropyfull{} is better than
\cwlossfull{}.''} Now, we want to understand the \textit{conditions} that make
a hypothesis true. These conditions can be using a specific dataset, under a
certain threat model, or based on other component values. Building off our
previous example, this hypothesis paired with a condition could be
\textit{``using \crossentropyfull{} is better than \cwlossfull{}, when the
dataset is \phishing{}.''} When we test a hypothesis, we look at the
statistical significance of the hypothesis under all conditions to
determine when a hypothesis is true. Enumerating across all
possible hypothesis and condition pairs yielded \num{1690} candidate hypotheses.
It should be noted that the component values in hypotheses are always in the
same component, as comparing usefulness across components would be nonsensical
(\eg{} ``Using \crossentropyfull{} is better than using \rr{}'' is not
meaningful).

\begin{table*}[!ht]
    \centering
    \resizebox{.9\textwidth}{!}{%
    \begin{tabular}{rcccclcc}
        \toprule
                                                &
        \textbf{Component}   \(\mathbf{H_1}\)   &
                                                &
        \textbf{Component}   \(\mathbf{H_2}\)   &
                                                &
        \multicolumn{1}{c}{\textbf{Condition}}  &
        \(\mathbf{p}\)-\textbf{value}           &
        \textbf{Effect Size}                    \\
        \midrule
1. & \sgd{} & is better than &  \bwsgd{} & when & \(\mathtt{Dataset}=\mnist{}\)
        & \num{< 2.2e-308} & \qty{99}{\percent} \\
2. & \adam{} & is better than &  \bwsgd{} & when &
        \(\mathtt{Dataset}=\mnist{}\) & \num{< 2.2e-308} & \qty{99}{\percent}
        \\
        \multicolumn{4}{c}{\vdots} & \multicolumn{4}{c}{\vdots} \\

84. & \identitylossfull{} & is better than &  \dlrlossfull{} & when &
        \(\mathtt{Dataset}=\nslkdd{}\) & \num{< 2.2e-308} & \qty{93}{\percent}
        \\
85. & \sgd{} & is better than &  \bwsgd{} & when & \(\mathtt{Saliency
        Map}=\jsmasmapfull{}\) & \num{< 2.2e-308} & \qty{92}{\percent} \\
        \multicolumn{4}{c}{\vdots} & \multicolumn{4}{c}{\vdots} \\

393. & \dfsmapfull{} & is better than &  \jsmasmapfull{} & when &
        \(\mathtt{Dataset}=\fmnist{}\) & \num{< 5e-6} & \qty{66}{\percent} \\
394. & \crossentropyfull{} & is better than &  \cwlossfull{} & when &
        \(\cov{}=\textit{Disabled}\) & \num{< 5e-6} & \qty{61}{\percent} \\
    \multicolumn{4}{c}{\vdots} & \multicolumn{4}{c}{\vdots} \\

1689. & \lp[0] & is better than &  \lp[2] & when & \(\mathtt{Threat\
        Model}=\lp[2]+1.0\) & \num{9.8e-01} & \qty{50}{\percent} \\
1690. & \identitysmapfull{} & is better than &  \dfsmapfull{} & when &
        \(\mathtt{Threat\ Model}=\lp[\infty]+0.4\) & \num{1.0e+00} &
        \qty{49}{\percent} \\
        \bottomrule
    \end{tabular}}
    \caption{The evaluated hypotheses for non-robust models. The top \num{344}
    hypotheses have a \(p\)-value that exhibits \qty{64}{\bit} underflow.
    When sorted by effect size, the top \qty{50}{\percent} of hypotheses have
    an effect size greater than \qty{80}{\percent}.}\label{tab:hypotheses}
\end{table*}

\subsubsection{Testing}

We test the \num{1690} hypotheses with the Wilcoxon Signed-Rank Test, a
non-parametric pairwise test, equivalent to a pairwise Mann-Whitney \(U\) Test,
to determine its significance. We also report the effect size of the test,
defined as the percentage of pairwise median areas (over ten trials, with trial
counts factored into computed \(p\)-values) from component \(H_1\) that were
smaller than component \(H_2\) (recall, a smaller area corresponds to a better
attack, as it more closely tracks the \oea{}). Note that the \(p\)-values for
many hypotheses underflowed \qty{64}{\bit} floating point precision, implying
that the results of the test are highly significant across all datasets and
threat models. A subset of of hypotheses are represented in
\autoref{tab:hypotheses}.

We find many highly-significant correlations in the results across the space of
hypotheses. Specifically, we set a significance threshold proportional to the
number of hypothesis tests we evaluated to minimize false
positives\footnotemark{}: \(p < \frac{0.01}{1690}=\num{5e-6}\). We found that
\num{1536} (\qty{90}{\percent}) of hypotheses were below this threshold. We
highlight the most prominent conclusions among these \num{1536} hypothesis:
\textbf{(1)} \cov{} was found to be disadvantageous---\num{86} hypotheses
involving \cov{} met our threshold; all \num{86} were against its use,
\textbf{(2)} \adam{} was superior to all other optimizers---\num{503}
hypotheses comparing \adam{} to other optimizers met our threshold, of which
\qty{50}{\percent} of them ruled in favor of \adam{} (with \sgd{} at
\qty{33}{\percent}, and \mbs{} at \qty{16}{\percent}), \textbf{(3)} \rr{} was
found to be preferable across \qty{61}{\percent} of hypotheses (\num{51} of
\num{83}), \textbf{(4)} \lp[\infty]-targeted attacks, at \qty{79}{\percent}
(\num{163} of \num{205}) were superior to both \lp[0]- and \lp[2]-targeted
(which were only favorable \qty{16}{\percent} (\num{34} of \num{205}) and
\qty{4}{\percent} (\num{8} of \num{205}) of the time, respectively),
\textbf{(5)} using no saliency map (\ie{} \identitysmap{}) was better
\qty{70}{\percent} (\num{131} of \num{187}) of the time, \textbf{(6)} perhaps
surprisingly, using no loss function was more advantageous \qty{47}{\percent}
(\num{224} of \num{472}) of the time, over \crossentropy{} and \cwloss{}, which
were useful \qty{34}{\percent} (\num{161} of \num{472}) and \qty{18}{\percent}
(\num{87} of \num{472}) of the time, respectively, and \textbf{(7)} contrary to
common practice, using \lp[\infty]-based attacks were sometimes superior to
\lp[2]-based attacks \textit{for \lp[2]-based threat models} (\num{21} of
\num{42}); this result would suggest that perturbing based on the magnitude of
gradients, while effective, can be excessive (when measuring cost under
\lp[2]) and unnecessary to meet adversarial goals.

\footnotetext{One would expect evaluating \num{1000} hypotheses at \(p<0.01\)
significance would result in \num{10} false positives, for example.}

We highlight some key takeaways from this experiment: (1) These hypothesis
tests provide statistical evidence of some common practices within
the community (using \rr{} and the superiority of \adam{}), while also
demonstrating some perhaps surprising conclusions, such as the detriment of
using \crossentropyfull{} over no loss function at all. (2) We emphasize the
utility of hypothesis testing for threat modeling as well: the tests provide a schema for performing worst-case benchmarks in their
respective domain. For example, when benchmarking \mnist{} against \lp[0]-based
adversaries, attacks that use the \jsmasmapfull{} are likely to outperform
attacks that use \dfsmapfull{}.

\subsubsection{The Effect of Model Robustness}\label{explainability:robust} As
shown in \autoref{fig:spearman_robust}, robust models can have a significant
impact on attack rankings. Here, we investigate why such broad phenomena occur.
Specifically, we investigate how attack parameter choices change performance on
a robust versus a non-robust model. We repeat our hypothesis testing on robust
models only and compare the hypotheses most affected (that is, the largest
changes in effect size) by robust models.

\autoref{tab:hypotheses_robust} provides a listing of the top pairs of
hypotheses, sorted by the change in effect size from a non-robust to robust
model (labeled as delta). Many of the top hypotheses when migrating from
non-robust to robust models largely concern \cifar{} and \malmem{}, which were
broadly the most unique phenomena across our experiments. Specifically, we see
large changes in losses and saliency maps for the attacks that were effective
at attacking robust models. The emphasis on \crossentropy{} could be in part
attributed to the fact that both the model is trained on this loss as well as
used by \pgd{}, the attack used to generate adversarial examples within
minibatches. This observation suggests that alignment between between attack
losses and losses used for adversarial training is highly effective at
attacking robust models.

Beyond the influence of loss on \cifar{} and \malmem{}, most of our tested
hypotheses remained relatively unaffected by model robustness: of the
\num{1690} hypotheses tested, only \num{334} had an effect size change of
\qty{10}{\percent} or greater between robust and non-robust models. This
implies that, while many of the factors that make attacks effective do not vary
between normally- and adversarially-trained models, the subset that
\textit{does} vary accounts for a vast difference in attack effectiveness.

\begin{table*}[!ht]
    \centering
    \resizebox{.9\textwidth}{!}{%
    \begin{tabular}{rcccclccc}
        \toprule
                                                &
        \textbf{Component}   \(\mathbf{H_1}\)   &
                                                &
        \textbf{Component}   \(\mathbf{H_2}\)   &
                                                &
        \multicolumn{1}{c}{\textbf{Condition}}  &
        \(\mathbf{p}\)-\textbf{value}           &
        \textbf{Effect Size}                    &
        \textbf{Delta}\\
        \midrule
1. & \crossentropyfull{} & is better than &  \dlrlossfull{} & when & \(\mathtt{Dataset}=\cifar{}\) & \num{< 2.2e-308} & \qty{96}{\percent} & \qty{45}{\percent} \\
2. & \identitysmapfull{} & is better than &  \dfsmapfull{} & when & \(\mathtt{Dataset}=\cifar{}\) & \num{< 2.2e-308} & \qty{74}{\percent} & \qty{44}{\percent} \\
3. & \dlrlossfull{} & is better than &  \cwlossfull{} & when & \(\mathtt{Dataset}=\nslkdd{}\) & \num{< 1e-5} & \qty{57}{\percent} & \qty{44}{\percent} \\
4. & \crossentropyfull{} & is better than &  \identitylossfull{} & when & \(\mathtt{Dataset}=\malmem{}\) & \num{< 2.2e-308} & \qty{69}{\percent} & \qty{43}{\percent} \\
5. & \rr{}: \textit{Disabled} & is better than &  \rr{}: \textit{Enabled} &
        when & \(\mathtt{Optimizer}=\bwsgd{}\) & \num{< 2.2e-308} & \qty{92}{\percent} & \qty{41}{\percent} \\
6. & \adam{} & is better than &  \sgd{} & when & \(\mathtt{Dataset}=\malmem{}\) & \num{< 1e-5} & \qty{46}{\percent} & \qty{39}{\percent} \\
7. & \rr{}: \textit{Disabled} & is better than &  \rr{}: \textit{Enabled} & when & \(\mathtt{Dataset}=\malmem{}\) & \num{< 2.2e-308} & \qty{90}{\percent} & \qty{35}{\percent} \\
8. & \identitylossfull{} & is better than &  \dlrlossfull{} & when & \(\mathtt{Dataset}=\nslkdd{}\) & \num{< 1e-5} & \qty{57}{\percent} & \qty{35}{\percent} \\
9. & \rr{}: \textit{Disabled} & is better than &  \rr{}: \textit{Enabled} & when & \(\mathtt{Dataset}=\unswnb{}\) & \num{< 1e-5} & \qty{65}{\percent} & \qty{33}{\percent} \\
10. & \cwlossfull{} & is better than &  \dlrlossfull{} & when & \(\mathtt{Dataset}=\cifar{}\) & \num{< 2.2e-308} & \qty{81}{\percent} & \qty{32}{\percent} \\
11. & \crossentropyfull{} & is better than &  \cwlossfull{} & when & \(\mathtt{Saliency Map}=\identitysmapfull{}\) & \num{< 2.2e-308} & \qty{83}{\percent} & \qty{31}{\percent} \\
12. & \crossentropyfull{} & is better than &  \dlrlossfull{} & when & \(\mathtt{Dataset}=\nslkdd{}\) & \num{6.4e-05} & \qty{55}{\percent} & \qty{31}{\percent} \\
13. & \crossentropyfull{} & is better than &  \identitylossfull{} & when & \(\mathtt{Saliency Map}=\identitysmapfull{}\) & \num{< 1e-5} & \qty{57}{\percent} & \qty{30}{\percent} \\
14. & \identitysmapfull{} & is better than &  \dfsmapfull{} & when & \(\mathtt{Loss}=\crossentropyfull{}\) & \num{< 2.2e-308} & \qty{79}{\percent} & \qty{30}{\percent} \\
15. & \identitysmapfull{} & is better than &  \jsmasmapfull{} & when & \(\mathtt{Dataset}=\cifar{}\) & \num{< 2.2e-308} & \qty{79}{\percent} & \qty{29}{\percent} \\
        \bottomrule
    \end{tabular}}
    \caption{The top 15 hypotheses for robust models. Delta represents the
    difference in effect size when changing to a robust
    model.}\label{tab:hypotheses_robust}
\end{table*}

\section{Discussion}\label{discussion}

\iflinks\hypertargetblue{mr5}{\mr{5}}\fi 

\shortsection{Domain Constraints} While adversarial machine learning research
has been cast predominantly through images, the threats imposed to
machine-learning-based detection systems via malware or network attacks are
increasingly concerning. However, producing legitimate adversarial examples in
the form of binaries or packet captures is a nuanced process; there are
constraints, dictated by the domain, that adversarial examples must comply
with~\cite{sheatsley_robustness_2021, chandrasekaran_rearchitecting_2019,
anderson_evading_nodate, yang_adversarial_2018, melacci_domain_2021, demetrio_practical_2022, melacci_domain_2021,travers_exploitability_2021,jain_analyzing_2019 }.

In addition, adversarial goals in such domains are not precisely captured by
\autoref{eq:goals}; attacks are commonly \textit{targeted} towards a specific
class (such as, classifying a variety of malicious network flows as benign
traffic~\cite{sheatsley_adversarial_2020, sheatsley_robustness_2021,
yang_adversarial_2018, lin_idsgan_2018} or malware families as legitimate
software~\cite{grosse_adversarial_2017, anderson_evading_nodate,
demetrio_functionality-preserving_2021, demetrio_practical_2022, kaur_robust_2022}). Moreover,
recent work has shown the unique challenges of producing adversarial examples
in the \textit{problem space}~\cite{demetrio_practical_2022, demetrio_functionality-preserving_2021, pierazzi_intriguing_2020}. Such works
identified a set of properties input perturbations must adhere to in order to be considered
demonstrative of malicious inputs in the respective problem space (\eg{} packet captures
or binaries), such as semantic preservation, problem-space transformations, robustness to preprocessing, among other important attributes.

These necessary factors provide a
more realistic perspective on the robustness of machine learning systems in
security-critical domains. While we did not explore these factors for scope, we
acknowledge their importance, and encourage subsequent investigations to
incorporate these factors (such as ensuring perturbations are
constraint-compliant at the \(\lp{}\) layer of surfaces or adapting loss
functions to ensure adversarial examples are misclassified as a specific target
class).

\shortsection{The Threat Landscape} \textit{White-box} adversaries are
important because they represent worst-case failure modes of machine learning
systems. However, \textit{black-box} adversaries have demonstrated remarkable
efficacy within their limited amount of available knowledge (\ie{} practical
threats)~\cite{papernot_practical_2017, ilyas_black-box_2018,
brendel_decision-based_2017, suya_hybrid_2020}. While this initial application
of our framework focused on white-box adversaries for their prevalence in
research, we note that there natural extensions to support black-box
adversaries, such as using \texttt{Backward Pass Differentiable
Approximation}~\cite{athalye_obfuscated_2018} in place of the model Jacobian,
or the Jacobian-based dataset augmentation~\cite{papernot_transferability_2016}
as a saliency map for training substitute models, among other techniques. As
there are a variety of techniques for efficiently mounting black-box attacks
(historically through query minimization)~\cite{papernot_practical_2017,
suya_hybrid_2020,shukla_simple_2021, cheng_sign-opt_2020}, we see value in
instantiating our framework with black-box components to understand the
trade-offs between such techniques.


\shortsection{Related Work} A natural limitation of \autoattackfull{} that the
ensemble is fixed; while it was designed to be as diverse as possible to common
failures of defenses, it may fail on defenses where an expert-designed adaptive
attack would succeed. Thus, the \texttt{Adaptive AutoAttack}
(\texttt{A}\textsuperscript{3}) extension was introduced to combine the
efficacy of \autoattackfull{}, while dynamically adapting to new
defenses~\cite{yao_automated_2021}. \texttt{A}\textsuperscript{3} frames
building adaptive attacks as a search problem, wherein a surrogate model is
built and a ``backbone'' attack (\eg{} \fgsm{}, \pgd{}, \cw{}, among others) is
greedily selected, paired with a loss function and subroutines (such as \rr{}).
\texttt{A}\textsuperscript{3} builds upon \autoattackfull{} in that it enables
searching through the attack design space to find the most effective adaptive
attack. Our work is complementary in that we provide a broad, modular attack
space, while \texttt{A}\textsuperscript{3} provides an approach for building
adaptive attacks dynamically.

\vspace{-2mm}
\section{Conclusion}\label{conclusion}

In this paper, we introduced the space of adversarial strategies. We first
presented an extensible decomposition of current attacks into their core
components. We subsequently constructed \num{568} previously unexplored attacks
by permuting these components. Through this vast attack space, we measured
attack optimality via the \oea{}: a theoretical attack that upper-bounds attack
performance. With the \oea{}, we studied how attack rankings change across
datasets, threat models, and robust vs non-robust models. From these rankings,
we described the space of hypotheses, wherein we evaluated how component
choices conditionally impact attack efficacy. Our investigation revealed that
attack performance is highly contextual---certain components can help (or hurt)
attack performance when a specific \lp{}-norm, compute budget, domain, and even
phenomena is considered. The space of adversarial strategies is rich with
highly competitive attacks; meaningful evaluations need to consider the myriad
of contextual factors that yield performant adversaries.

\section{Acknowledgements}
We would like to thank the anonymous reviewers for their insightful 
feedback throughout the review cycle. Additionally, we would like
to thank Quinn Burke, Yohan Beugin, Rachel King, and Dave Evans for their helpful comments
on earlier versions of this paper.

This research was sponsored by the Combat Capabilities Development Command Army Research Laboratory and was accomplished under Cooperative Agreement Number W911NF-13-2-0045 (ARL Cyber Security CRA). The views and conclusions contained in this document are those of the authors and should not be interpreted as representing the official policies, either expressed or implied, of the Combat Capabilities Development Command Army Research Laboratory or the U.S. Government. The U.S. Government is authorized to reproduce and distribute reprints for Government purposes not withstanding any copyright notation here on. This material is based upon work supported by the National Science Foundation under Grant No. CNS-1805310 and the U.S. Army Research Laboratory and the U.S. Army Research Office under Grant No. W911NF-19-1-0374.
\bibliographystyle{plain}
\bibliography{references_copy}
\ifarxiv
    \clearpage\appendix\onecolumn\section{Attack Encoding}\label{appendix-b:encoding}

Here we provide \autoref{tab:translation} for translating attack numbers to
component values.

\begin{table*}[ht!]
    \centering
    \resizebox{.9\textwidth}{!}{%
        \notsotiny{}
    \begin{minipage}{0.35\textwidth}
\begin{tabular}{lllllll}
        \toprule
        \textbf{\fw[\#]} & \textbf{Opt.} & \textbf{CoV}& \textbf{RR} & \textbf{\lp{}} &
        \textbf{SM} &  \textbf{Loss} \\
        \midrule
        0 & \sgd{} & \textit{False} & \textit{False} & \lp[0] & \identitysmap{} & \identityloss{}\\1 & \sgd{} & \textit{False} & \textit{False} & \lp[2] & \identitysmap{} & \identityloss{}\\2 & \sgd{} & \textit{False} & \textit{False} & \lp[\infty] & \identitysmap{} & \identityloss{}\\\texttt{JSMA} & \sgd{} & \textit{False} & \textit{False} & \lp[0] & \jsmasmap{} & \identityloss{}\\4 & \sgd{} & \textit{False} & \textit{False} & \lp[2] & \jsmasmap{} & \identityloss{}\\5 & \sgd{} & \textit{False} & \textit{False} & \lp[\infty] & \jsmasmap{} & \identityloss{}\\6 & \sgd{} & \textit{False} & \textit{False} & \lp[0] & \dfsmap{} & \identityloss{}\\
\bottomrule
        \end{tabular}
    \end{minipage}\hfill
    \begin{minipage}{0.35\textwidth}
        \begin{tabular}{lllllll}
        \toprule
        \textbf{\fw[\#]} & \textbf{Opt.} & \textbf{CoV}& \textbf{RR} & \textbf{\lp{}} &
        \textbf{SM} &  \textbf{Loss} \\
        \midrule
        96 & \sgd{} & \textit{False} & \textit{True} & \lp[0] & \dfsmap{} & \cwloss{}\\97 & \sgd{} & \textit{False} & \textit{True} & \lp[2] & \dfsmap{} & \cwloss{}\\98 & \sgd{} & \textit{False} & \textit{True} & \lp[\infty] & \dfsmap{} & \cwloss{}\\99 & \sgd{} & \textit{False} & \textit{True} & \lp[0] & \identitysmap{} & \dlrloss{}\\100 & \sgd{} & \textit{False} & \textit{True} & \lp[2] & \identitysmap{} & \dlrloss{}\\101 & \sgd{} & \textit{False} & \textit{True} & \lp[\infty] & \identitysmap{} & \dlrloss{}\\102 & \sgd{} & \textit{False} & \textit{True} & \lp[0] & \jsmasmap{} & \dlrloss{}\\
\bottomrule
        \end{tabular}
    \end{minipage}\hfill
    \begin{minipage}{0.35\textwidth}
        \begin{tabular}{lllllll}
        \toprule
        \textbf{\fw[\#]} & \textbf{Opt.} & \textbf{CoV}& \textbf{RR} & \textbf{\lp{}} &
        \textbf{SM} &  \textbf{Loss} \\
        \midrule
        192 & \adam{} & \textit{True} & \textit{False} & \lp[0] & \jsmasmap{} & \crossentropy{}\\193 & \adam{} & \textit{True} & \textit{False} & \lp[2] & \jsmasmap{} & \crossentropy{}\\194 & \adam{} & \textit{True} & \textit{False} & \lp[\infty] & \jsmasmap{} & \crossentropy{}\\195 & \adam{} & \textit{True} & \textit{False} & \lp[0] & \dfsmap{} & \crossentropy{}\\196 & \adam{} & \textit{True} & \textit{False} & \lp[2] & \dfsmap{} & \crossentropy{}\\197 & \adam{} & \textit{True} & \textit{False} & \lp[\infty] & \dfsmap{} & \crossentropy{}\\198 & \adam{} & \textit{True} & \textit{False} & \lp[0] & \identitysmap{} & \cwloss{}\\
\bottomrule
        \end{tabular}
    \end{minipage}}
\end{table*}

\begin{table*}[ht!]
    \centering
    \resizebox{.9\textwidth}{!}{%
        \notsotiny{}
    \begin{minipage}{0.35\textwidth}
        \begin{tabular}{lllllll}
        \toprule
        \textbf{\fw[\#]} & \textbf{Opt.} & \textbf{CoV}& \textbf{RR} & \textbf{\lp{}} &
        \textbf{SM} &  \textbf{Loss} \\
        \midrule
        288 & \bwsgd{} & \textit{False} & \textit{False} & \lp[0] & \identitysmap{} & \identityloss{}\\289 & \bwsgd{} & \textit{False} & \textit{False} & \lp[2] & \identitysmap{} & \identityloss{}\\290 & \bwsgd{} & \textit{False} & \textit{False} & \lp[\infty] & \identitysmap{} & \identityloss{}\\291 & \bwsgd{} & \textit{False} & \textit{False} & \lp[0] & \jsmasmap{} & \identityloss{}\\292 & \bwsgd{} & \textit{False} & \textit{False} & \lp[2] & \jsmasmap{} & \identityloss{}\\293 & \bwsgd{} & \textit{False} & \textit{False} & \lp[\infty] & \jsmasmap{} & \identityloss{}\\294 & \bwsgd{} & \textit{False} & \textit{False} & \lp[0] & \dfsmap{} & \identityloss{}\\
\bottomrule
        \end{tabular}
    \end{minipage}\hfill
    \begin{minipage}{0.35\textwidth}
        \begin{tabular}{lllllll}
        \toprule
        \textbf{\fw[\#]} & \textbf{Opt.} & \textbf{CoV}& \textbf{RR} & \textbf{\lp{}} &
        \textbf{SM} &  \textbf{Loss} \\
        \midrule
        384 & \bwsgd{} & \textit{False} & \textit{True} & \lp[0] & \dfsmap{} & \cwloss{}\\385 & \bwsgd{} & \textit{False} & \textit{True} & \lp[2] & \dfsmap{} & \cwloss{}\\386 & \bwsgd{} & \textit{False} & \textit{True} & \lp[\infty] & \dfsmap{} & \cwloss{}\\387 & \bwsgd{} & \textit{False} & \textit{True} & \lp[0] & \identitysmap{} & \dlrloss{}\\388 & \bwsgd{} & \textit{False} & \textit{True} & \lp[2] & \identitysmap{} & \dlrloss{}\\389 & \bwsgd{} & \textit{False} & \textit{True} & \lp[\infty] & \identitysmap{} & \dlrloss{}\\390 & \bwsgd{} & \textit{False} & \textit{True} & \lp[0] & \jsmasmap{} & \dlrloss{}\\
\bottomrule
        \end{tabular}
    \end{minipage}\hfill
    \begin{minipage}{0.35\textwidth}
        \begin{tabular}{lllllll}
        \toprule
        \textbf{\fw[\#]} & \textbf{Opt.} & \textbf{CoV}& \textbf{RR} & \textbf{\lp{}} &
        \textbf{SM} &  \textbf{Loss} \\
        \midrule
        480 & \mbs{} & \textit{True} & \textit{False} & \lp[0] & \jsmasmap{} & \crossentropy{}\\481 & \mbs{} & \textit{True} & \textit{False} & \lp[2] & \jsmasmap{} & \crossentropy{}\\482 & \mbs{} & \textit{True} & \textit{False} & \lp[\infty] & \jsmasmap{} & \crossentropy{}\\483 & \mbs{} & \textit{True} & \textit{False} & \lp[0] & \dfsmap{} & \crossentropy{}\\484 & \mbs{} & \textit{True} & \textit{False} & \lp[2] & \dfsmap{} & \crossentropy{}\\485 & \mbs{} & \textit{True} & \textit{False} & \lp[\infty] & \dfsmap{} & \crossentropy{}\\486 & \mbs{} & \textit{True} & \textit{False} & \lp[0] & \identitysmap{} & \cwloss{}\\
\bottomrule
        \end{tabular}
    \end{minipage}}
    \caption{Attack Name Encodings.}\label{tab:translation}
\end{table*}

    \newpage\onecolumn\section{Comparison to Known Attacks}\label{appendix-c}
\iflinks\hypertargetblue{mr6c}{\mr{6c}}\fi Here we provide the full version of \autoref{tab:mnist_comparison}.
\begin{table*}[!th]
    \centering
    \resizebox{.95\textwidth}{!}{%
    \tiny
    \begin{minipage}{0.52\textwidth}
    \setlength{\tabcolsep}{0.25em}
    \begin{tabular}{rccc rccc rccc}
        \toprule
        \multicolumn{4}{c}{\textbf{\lp[0] Attacks}} &
        \multicolumn{4}{c}{\textbf{\lp[2] Attacks}} &
        \multicolumn{4}{c}{\textbf{\lp[\infty] Attacks}} \\
        \cmidrule(lr){1-4}\cmidrule(lr){5-8}\cmidrule(lr){9-12}
        Rank & Attack & \% Reduction & \lp[0] Budget      &
        Rank & Attack & \% Reduction & \lp[2] Budget      &
        Rank & Attack & \% Reduction & \lp[\infty] Budget \\
        \midrule
        1.   & \fw[171] & -\qty{41}{\percent} & 0.10 &
        1.   & \fw[460] & -\qty{50}{\percent} & 0.12 &
        1.   & \fw[449] & -\qty{8}{\percent} & 0.22 \\
        2.   & \fw[27]  & -\qty{41}{\percent} & 0.10 &
        2.   & \fw[532] & -\qty{50}{\percent} & 0.12 &
        2.   & \fw[440] & -\qty{8}{\percent} & 0.22 \\
        3.   & \fw[459] & -\qty{41}{\percent} & 0.10 &
        3.   & \fw[454] & -\qty{45}{\percent} & 0.13 &
        3.   & \fw[17]  & -\qty{8}{\percent} & 0.22 \\
        4.   & \fw[144] & -\qty{41}{\percent} & 0.10 &
        4.   & \fw[526] & -\qty{45}{\percent} & 0.13 &
        4.   & \fw[8]   & -\qty{8}{\percent} & 0.22 \\
        5.   & \fw[150] & -\qty{41}{\percent} & 0.10 &
        5.   & \fw[451] & -\qty{41}{\percent} & 0.14 &
        5.   & \fw[458] & -\qty{4}{\percent} & 0.23 \\
        6.   & \fw[159] & -\qty{41}{\percent} & 0.10 &
        6.   & \fw[523] & -\qty{41}{\percent} & 0.14 &
        6.   & \fw[512] & -\qty{4}{\percent} & 0.23 \\
\bottomrule
    \end{tabular}
    \end{minipage}\hfill
    \begin{minipage}{0.52\textwidth}
    \setlength{\tabcolsep}{0.25em}
    \begin{tabular}{rccc rccc rccc}
        \toprule
        \multicolumn{4}{c}{\textbf{\lp[0] Attacks}} &
        \multicolumn{4}{c}{\textbf{\lp[2] Attacks}} &
        \multicolumn{4}{c}{\textbf{\lp[\infty] Attacks}} \\
        \cmidrule(lr){1-4}\cmidrule(lr){5-8}\cmidrule(lr){9-12}
        Rank & Attack & \% Reduction & \lp[0] Budget      &
        Rank & Attack & \% Reduction & \lp[2] Budget      &
        Rank & Attack & \% Reduction & \lp[\infty] Budget \\
        \midrule
        97.  & \fw[405] & \(+\infty\%\) & \(\infty\) &
        97.  & \fw[427] & \(+\infty\%\) & \(\infty\) &
        97.  & \fw[545] & +\qty{237}{\percent} & 0.81 \\
        98.  & \fw[408] & \(+\infty\%\) & \(\infty\) &
        98.  & \fw[403] & \(+\infty\%\) & \(\infty\) &
        98.  & \fw[113] & +\qty{237}{\percent} & 0.81 \\
        99.  & \fw[429] & \(+\infty\%\) & \(\infty\) &
        99.  & \fw[397] & \(+\infty\%\) & \(\infty\) &
        99.  & \fw[209] & +\qty{237}{\percent} & 0.81 \\
        100. & \fw[423] & \(+\infty\%\) & \(\infty\) &
        100. & \fw[400] & \(+\infty\%\) & \(\infty\) &
        100. & \fw[560] & +\qty{241}{\percent} & 0.82 \\
        101. & \fw[426] & \(+\infty\%\) & \(\infty\) &
        101. & \fw[385] & \(+\infty\%\) & \(\infty\) &
        101. & \fw[272] & +\qty{241}{\percent} & 0.82 \\
        102. & \fw[402] & \(+\infty\%\) & \(\infty\) &
        102. & \fw[379] & \(+\infty\%\) & \(\infty\) &
        102. & \fw[128] & +\qty{241}{\percent} & 0.82 \\
\bottomrule
    \end{tabular}
    \end{minipage}}
    \caption{\mnist{} Relative Attack Comparisons. Budgets are normalized.
    Attacks that fail to reduce model accuracy to be \qty{<1}{\percent} are
    labeled as consuming infinite budget. Improvements are relative to the best
    known attack for each \lp{}-norm.}
\end{table*}

    \newpage\twocolumn\section{The Optimal Attack}\label{appendix-d}

Here, we show the median Spearman rank correlation coefficients for each of our
datasets and threat models. For the dataset plots, results are segmented by
\lp{}-norm. Data points correspond to a specific threat model (\ie{} a value
for \(\theta\)). \(\theta=\set{0.0, 0.5, 1.0, 1.5, 2.0}\) are labeled for
reference. High attack performance generalization is encoded as lighter shades,
while low generalization is encoded with darker shades, as shown in the colobar
below:

\begin{figure}[!h]
    \centering
    \resizebox{!}{0.4in}{\includegraphics{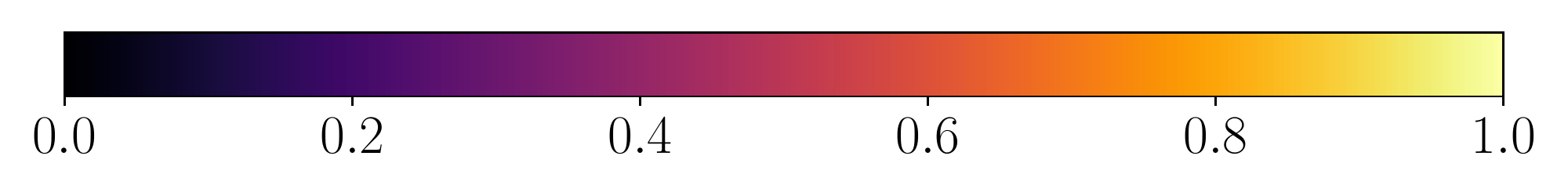}}
\end{figure}

\begin{figure}[H]
    \centering
    \resizebox{0.9\columnwidth}{!}{\includegraphics{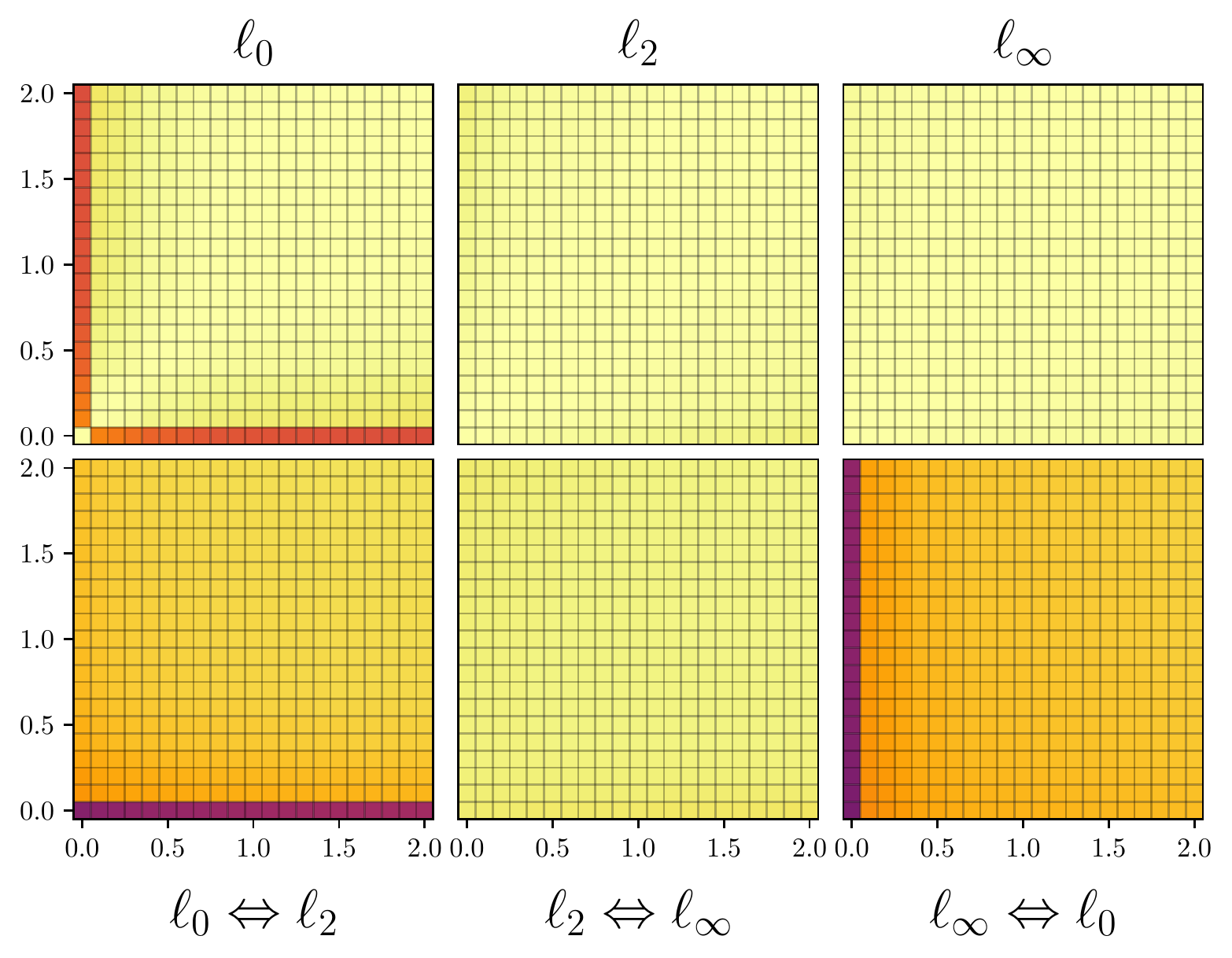}}
    \caption{\phishing}
\end{figure}

\begin{figure}[H]
    \centering
    \resizebox{0.9\columnwidth}{!}{\includegraphics{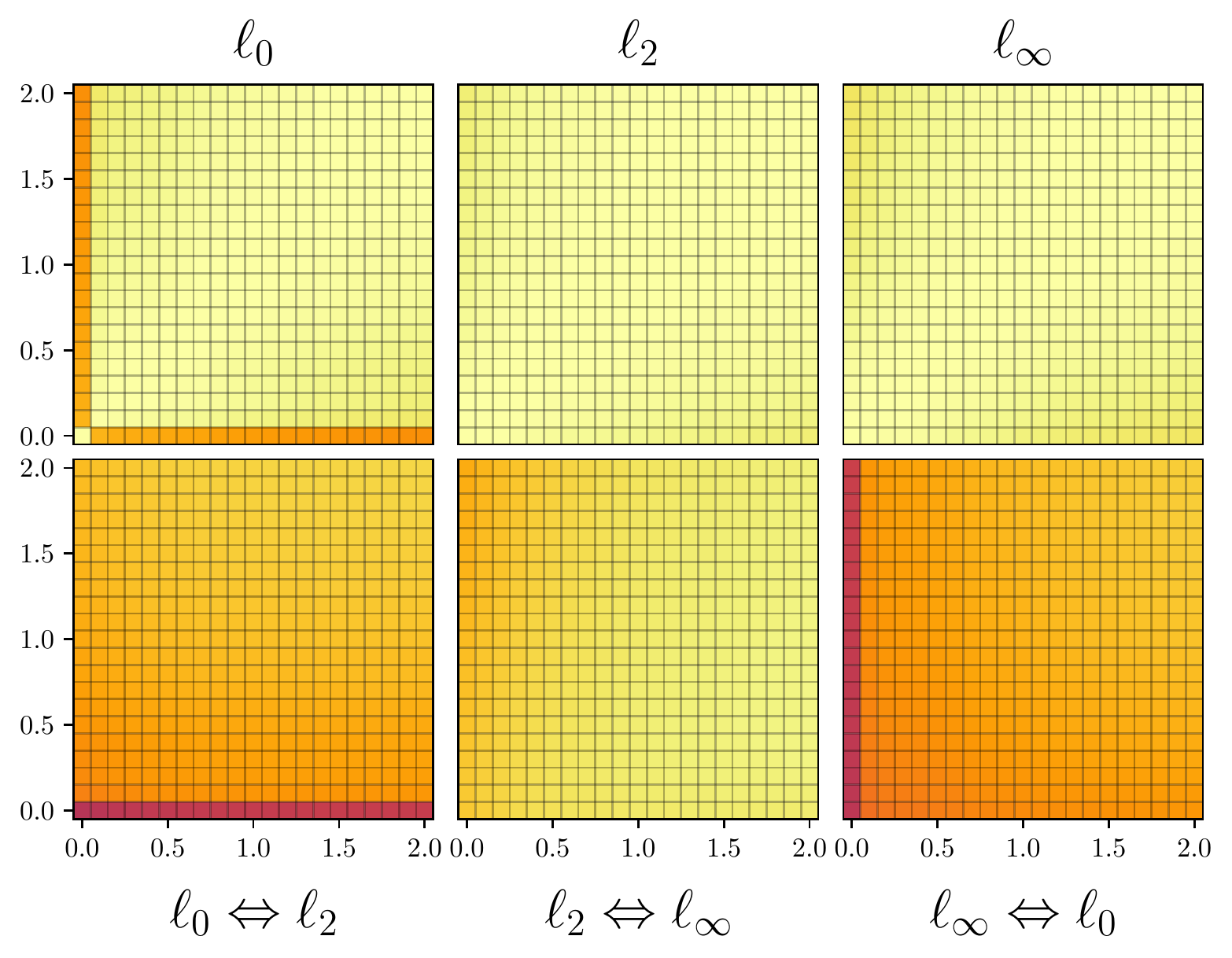}}
    \caption{\nslkdd}
\end{figure}

\begin{figure}[H]
    \centering
    \resizebox{0.9\columnwidth}{!}{\includegraphics{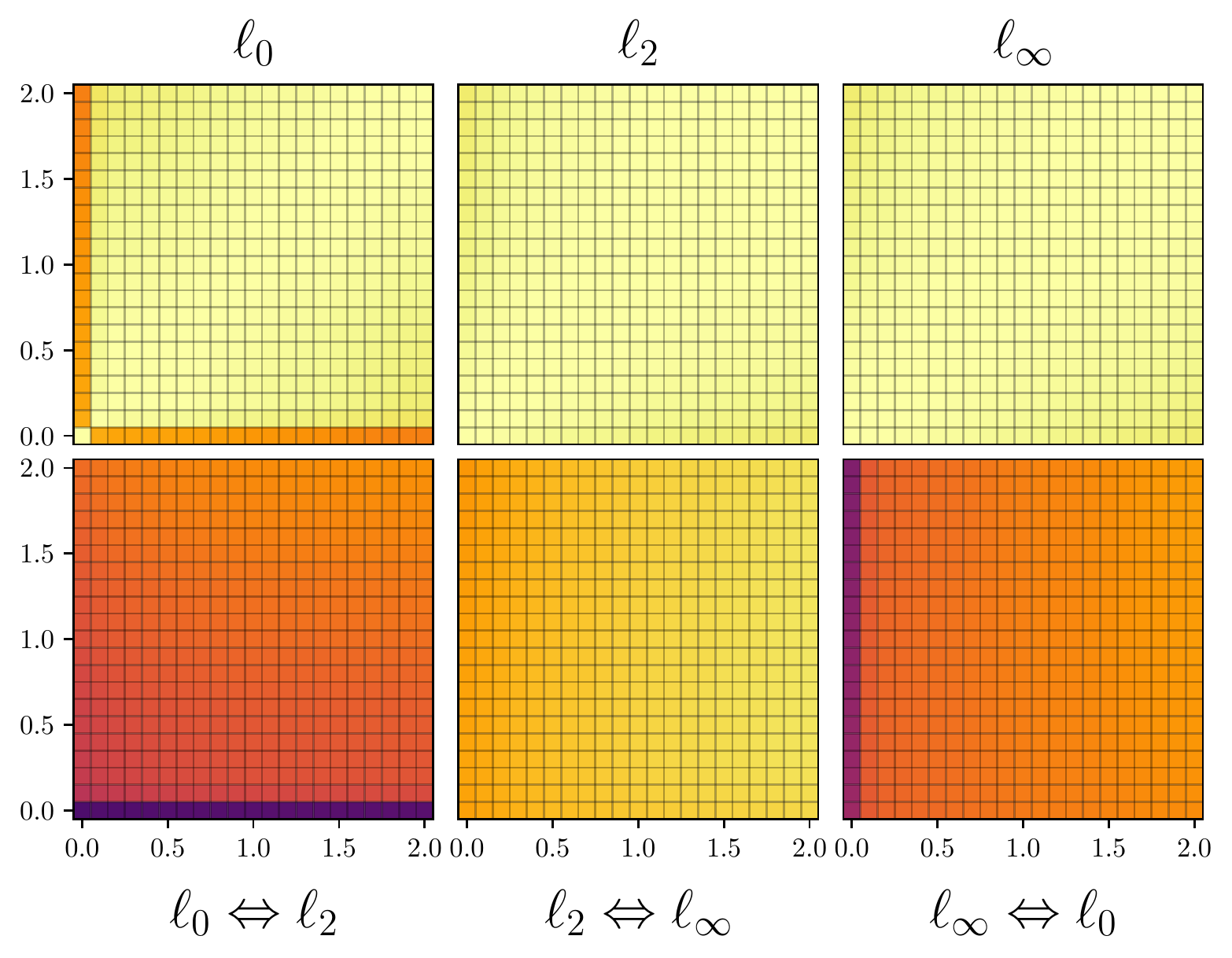}}
    \caption{\unswnb}
\end{figure}

\begin{figure}[H]
    \centering
    \resizebox{0.9\columnwidth}{!}{\includegraphics{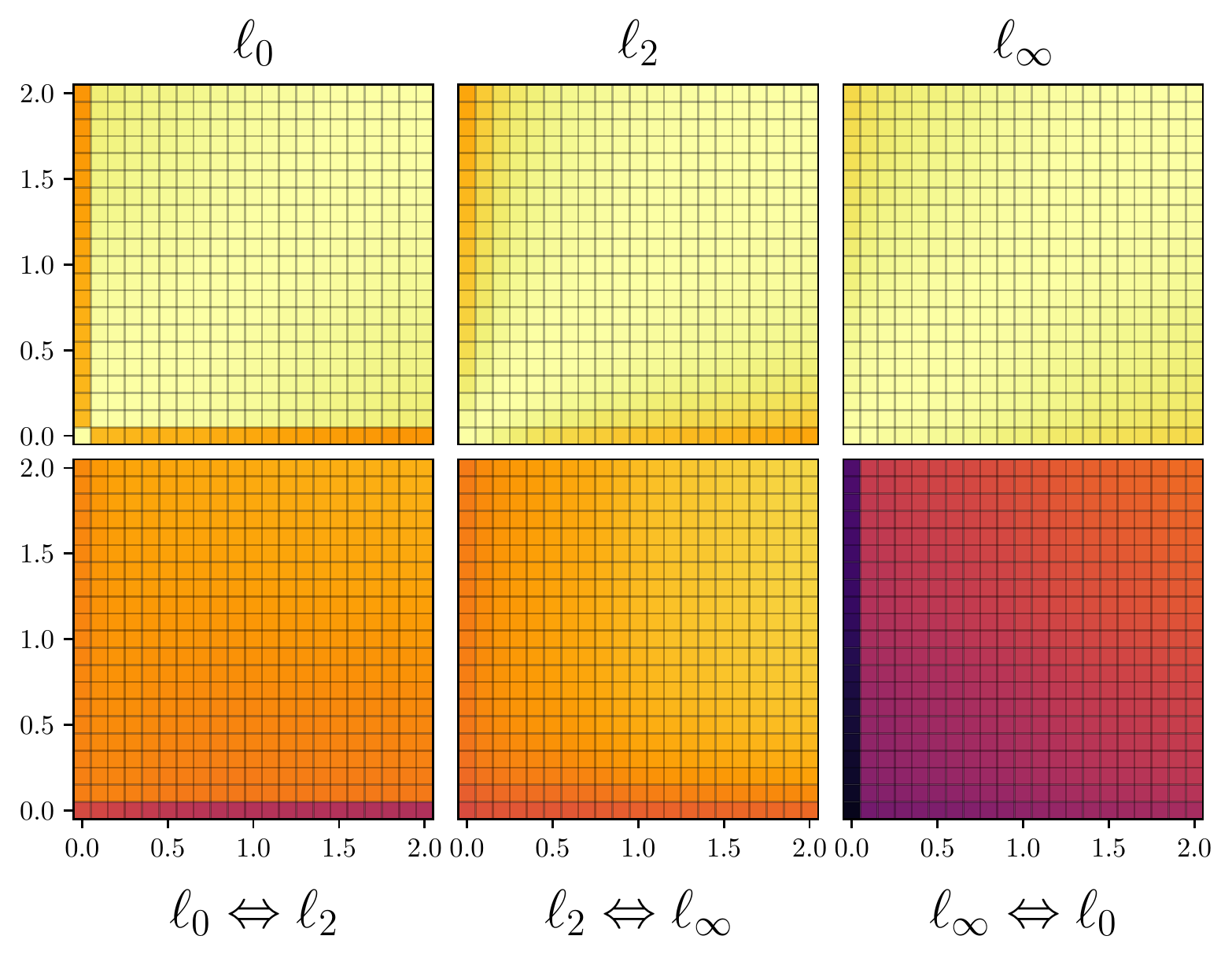}}
    \caption{\mnist}
\end{figure}

\begin{figure}[H]
    \centering
    \resizebox{0.9\columnwidth}{!}{\includegraphics{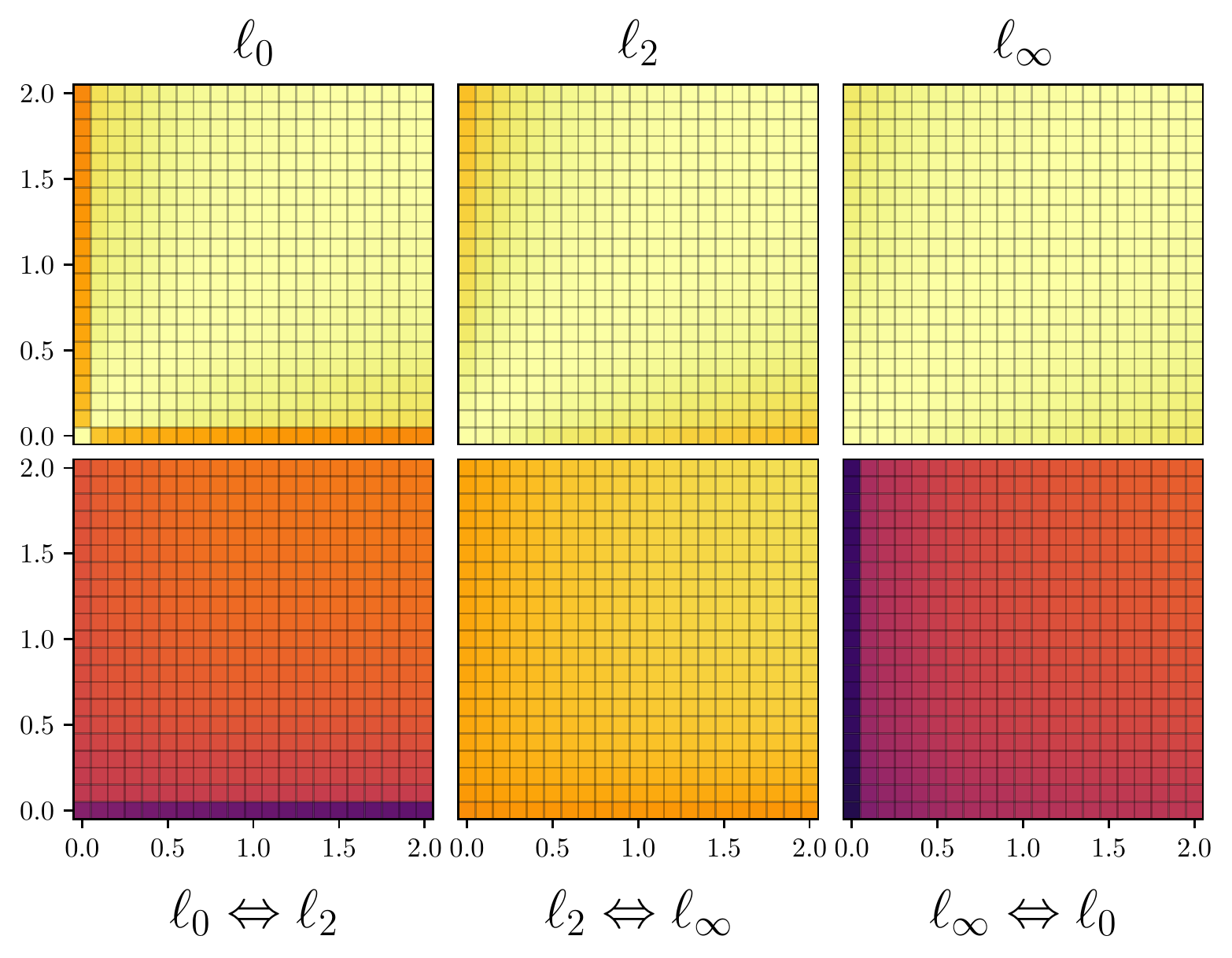}}
    \caption{\fmnist}
\end{figure}
\begin{figure}[H]
    \centering
    \resizebox{0.8\columnwidth}{!}{\includegraphics{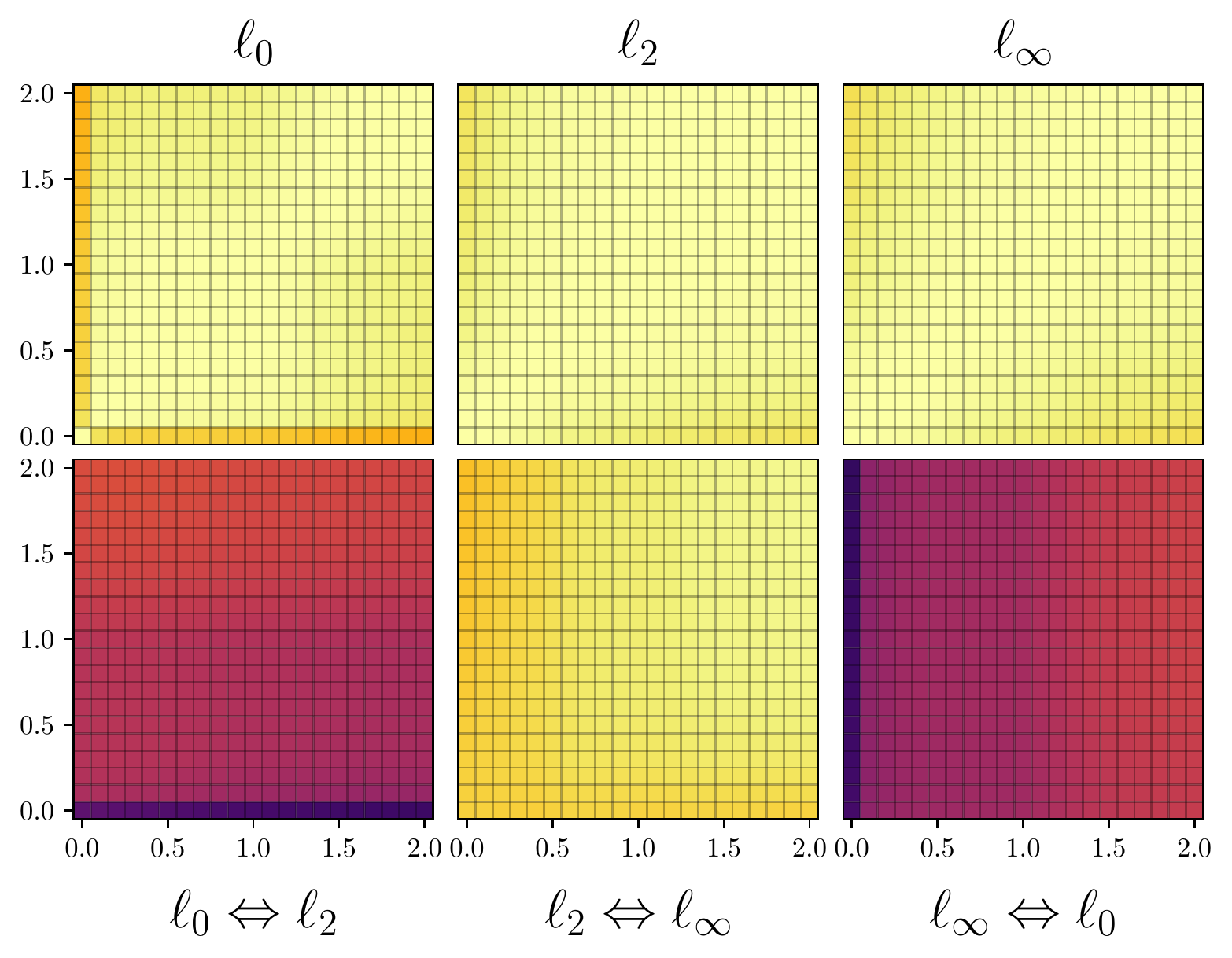}}
    \caption{\cifar{}}
\end{figure}

\begin{figure}[H]
    \centering
    \resizebox{0.8\columnwidth}{!}{\includegraphics{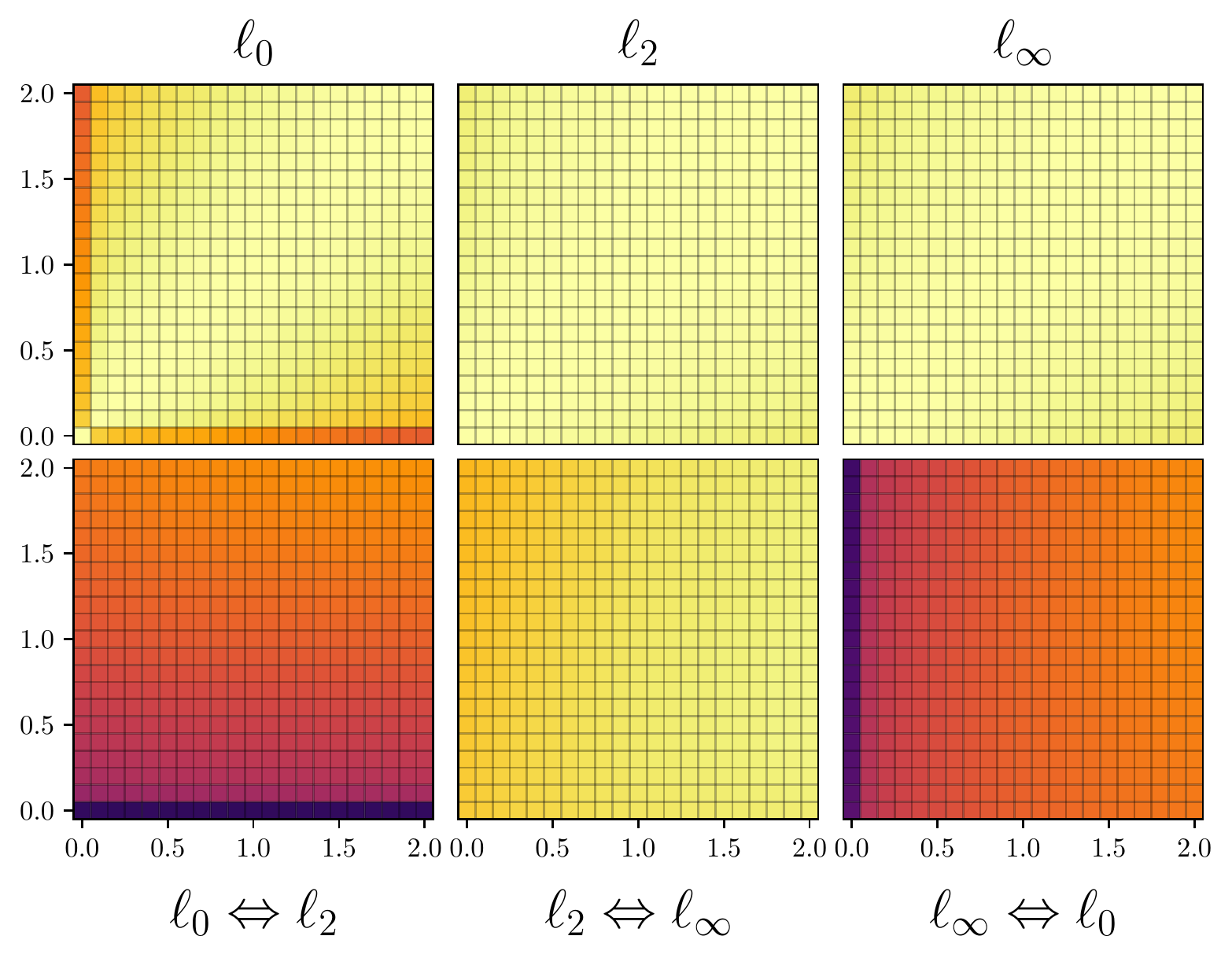}}
    \caption{\malmemfull{}}
\end{figure}
\begin{figure}[H]
    \centering
    \resizebox{0.49\textwidth}{!}{\includegraphics{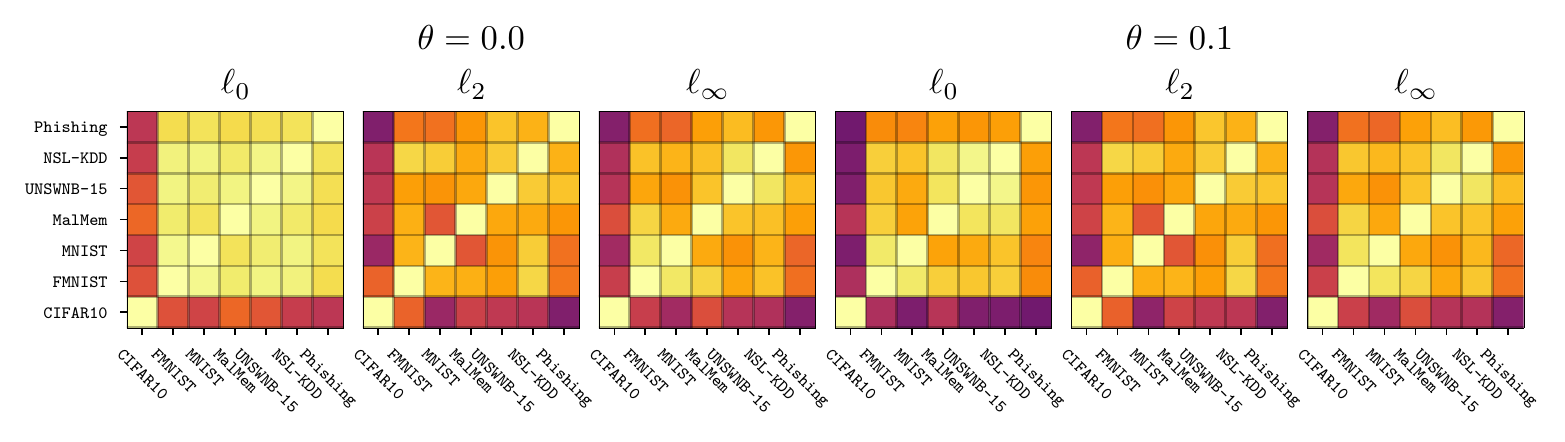}}
    \resizebox{0.49\textwidth}{!}{\includegraphics{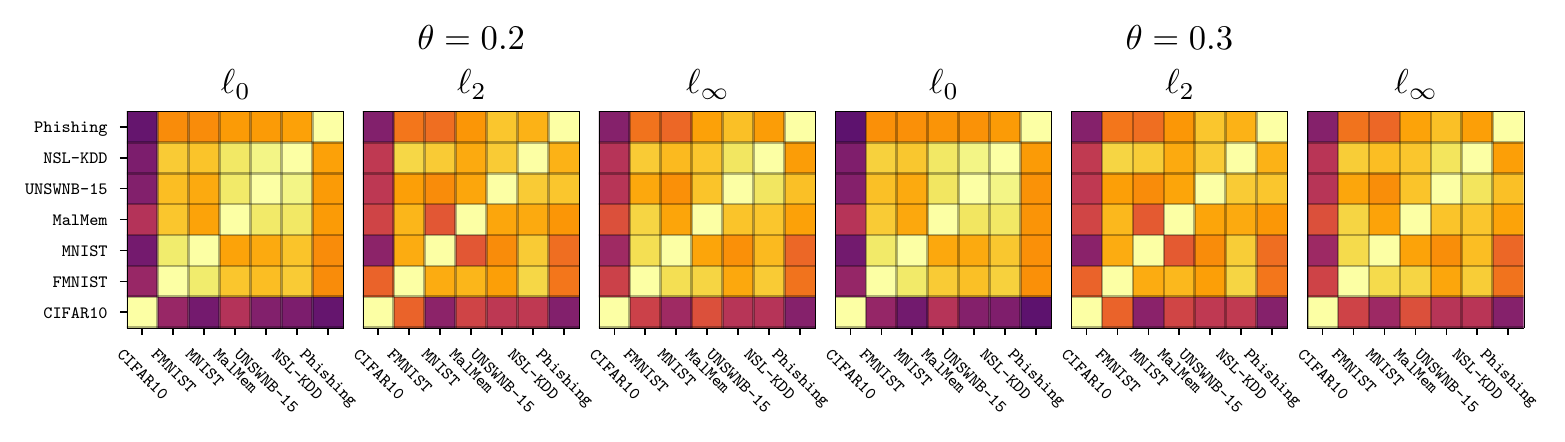}}
\end{figure}
\begin{figure}[H]
    \centering
    \resizebox{0.49\textwidth}{!}{\includegraphics{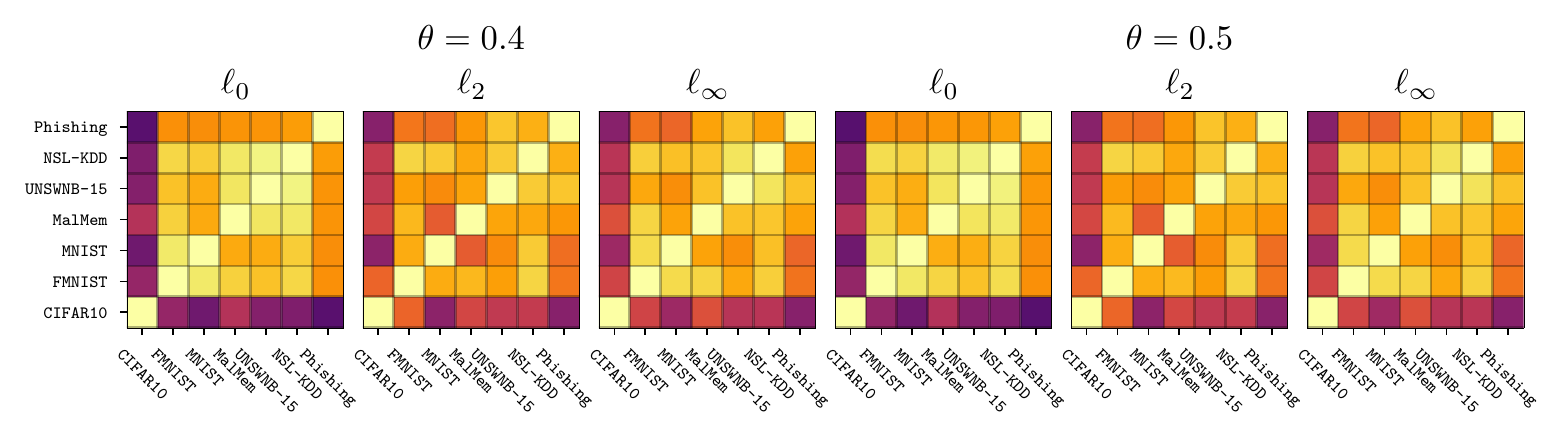}}
    \resizebox{0.49\textwidth}{!}{\includegraphics{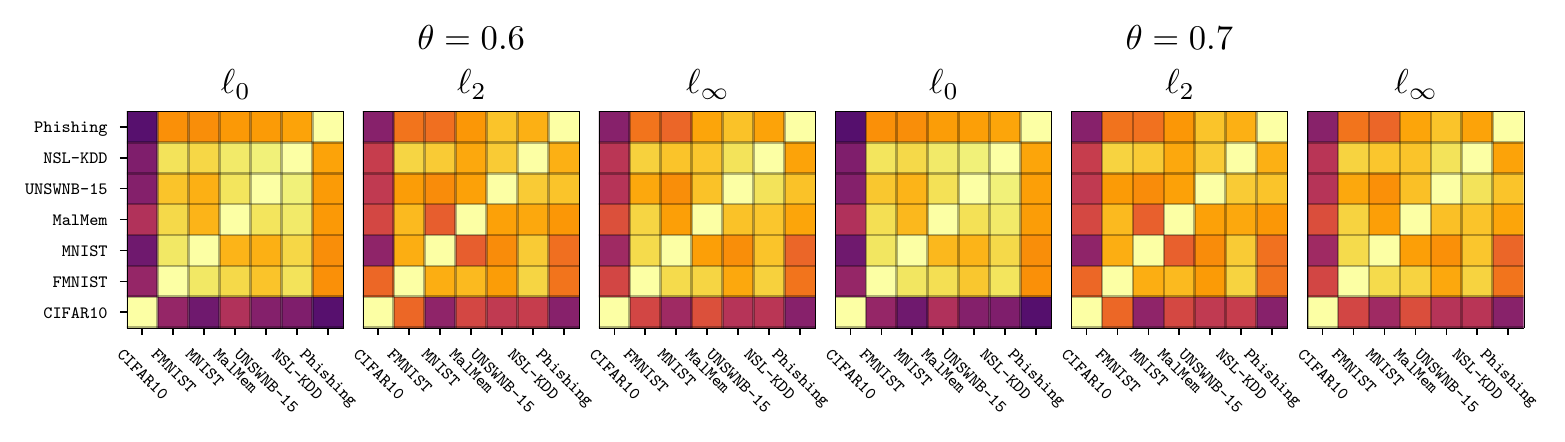}}
    \resizebox{0.49\textwidth}{!}{\includegraphics{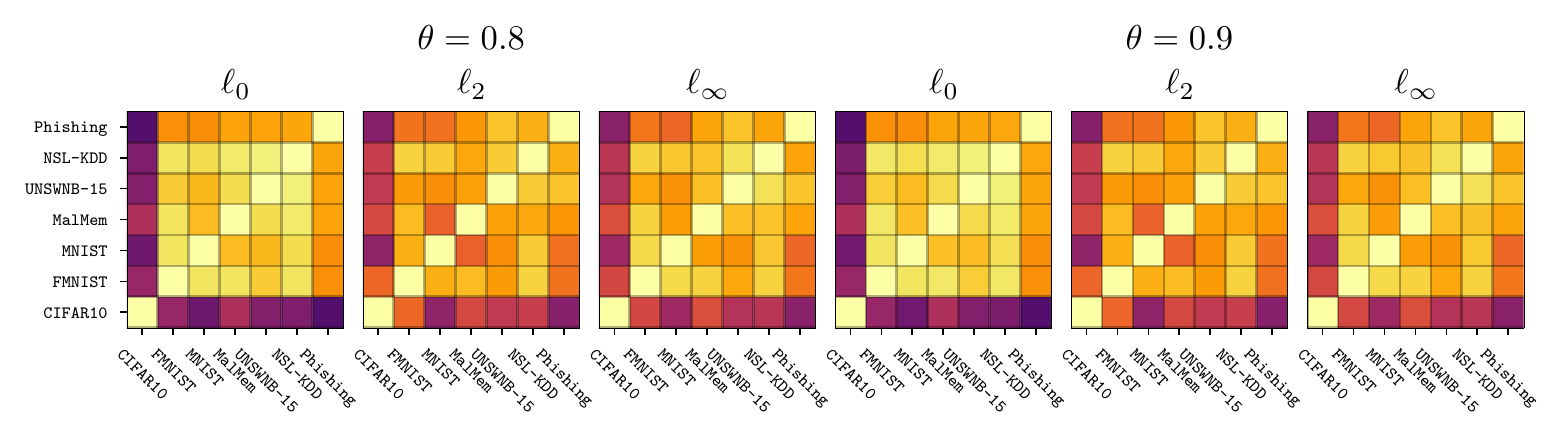}}
    \resizebox{0.49\textwidth}{!}{\includegraphics{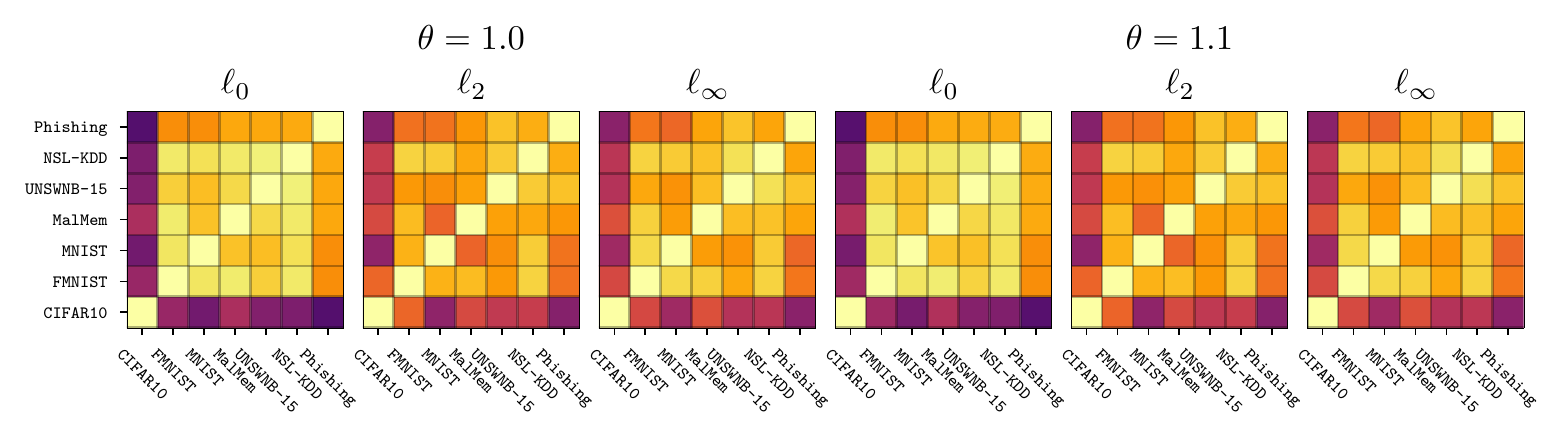}}
    \resizebox{0.49\textwidth}{!}{\includegraphics{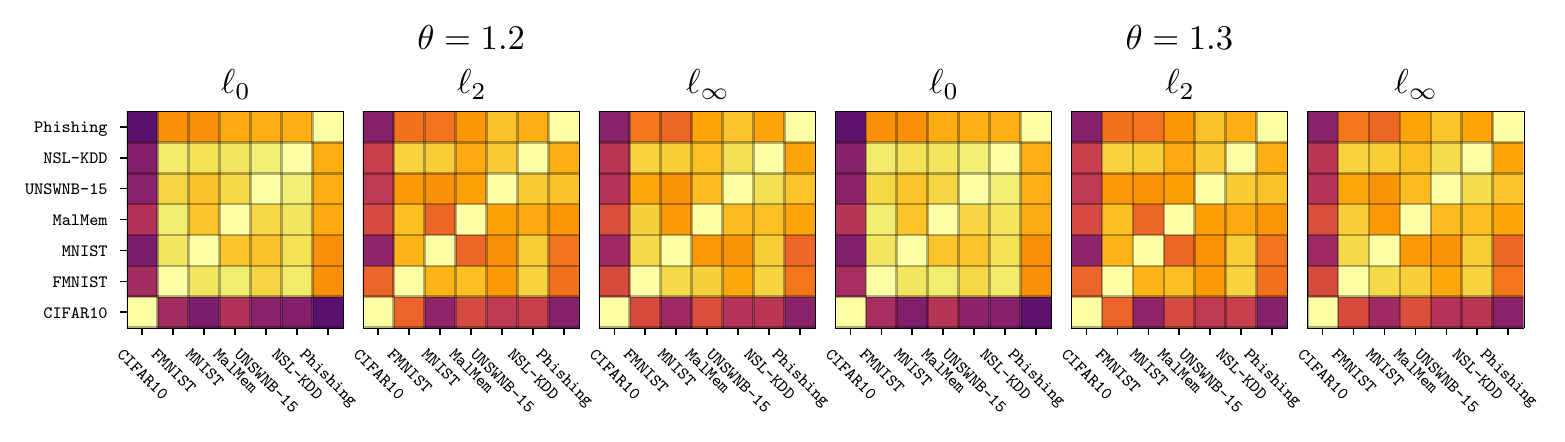}}
    \resizebox{0.49\textwidth}{!}{\includegraphics{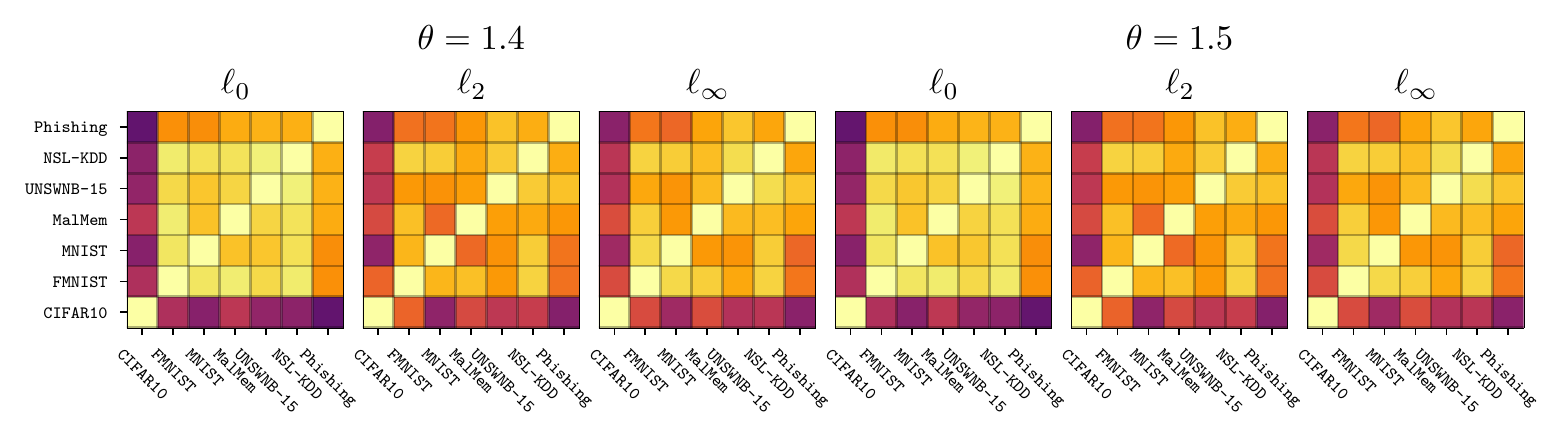}}
    \resizebox{0.49\textwidth}{!}{\includegraphics{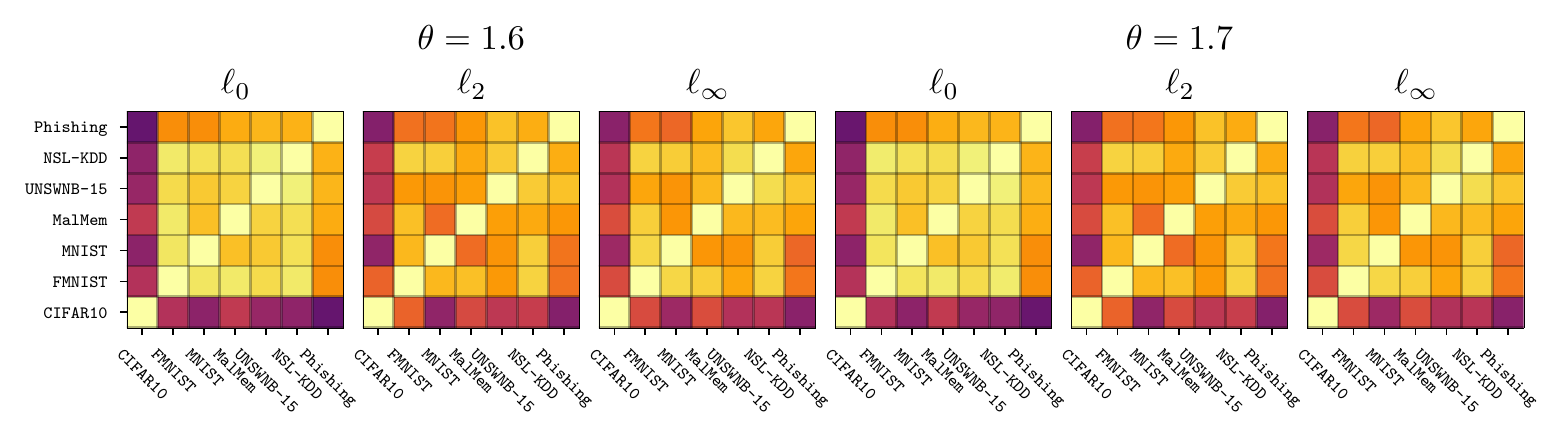}}
    \resizebox{0.49\textwidth}{!}{\includegraphics{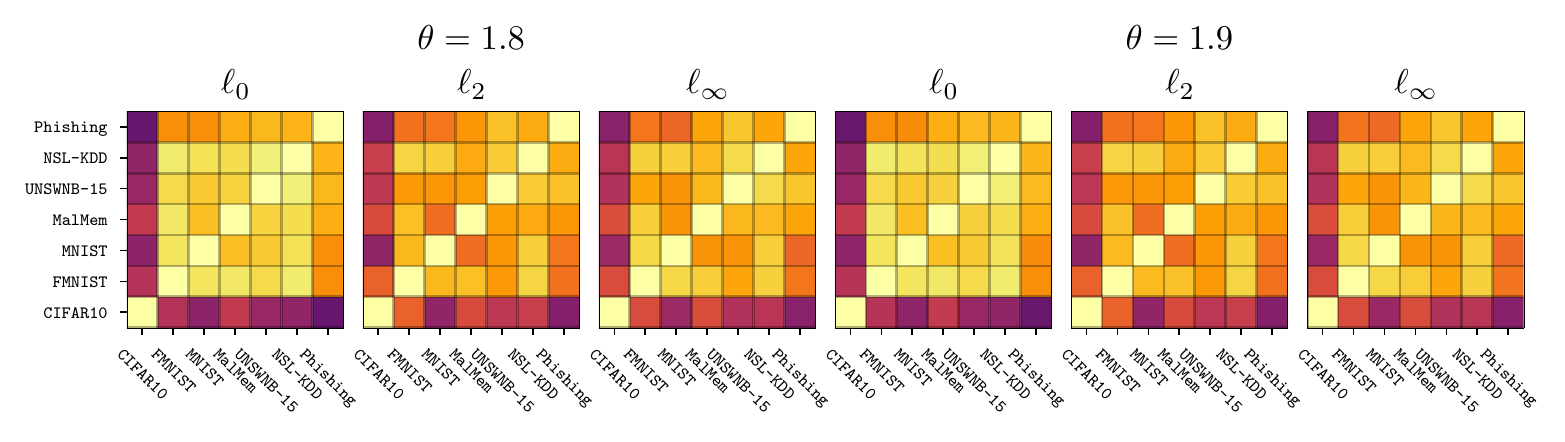}}
\end{figure}
\twocolumn
\fi
\newpage\twocolumn\ifarxiv\else\appendix\fi\section{Miscellany}\label{appendix-e}

\autoref{tab:hyperparameters} provides a listing of model hyperparameters for
each of our datasets. Our selection was inspired by publications that report
state-of-the-art accuracy for the models we used. \autoref{tab:symbols}
provides a listing of all symbols used in this paper and their associated
meanings. \autoref{tab:adv_hyperparameters} provides the parameters used for
adversarial training. \ifarxiv\else Finally, we provide \autoref{tab:translated} for translating attack numbers to
component values.\fi

\subsection{Attack Modifications}\label{appendix-e:modifications}

\shortsection{\cwfull{}} As described in \autoref{decomposition}, the
\cw{} attack loss function includes a hyperparameter \(c\) which controls the
trade-off between the distortion introduced and misclassification. In the
original attack definition, \(c\) is optimized dynamically through
binary-search~\cite{carlini_towards_2017}. This is cost-prohibitive and
prevents us from performing any meaningful evaluation when computational cost
is considered (as this attack would exist on a separate scale, when compared to
\pgd{} or even the \jsma{}, which requires the model Jacobian). To remedy this,
we select a constant value of \(c\) in our experiments. From the investigation
on values of \(c\) in~\cite{carlini_towards_2017} with respect to attack
success probability versus mean \lp[2] distance, we choose a value of \num{1.0}
for \(c\) in all experiments.

\shortsection{\jsmafull{}} The original definition of the \jsma{} included a
\textit{search space}, which defined the set of candidate features to be
selected for perturbation. In the original publication, the \jsma{} initially
set \(\alpha\) to either \num{1} or \num{0} (that is, pixels were fully turned
``off'' or ``on''). We find that this underestimates the performance
of the \jsma{} on many datasets. Instead, we derive a more effective strategy
of instead setting the saliency map score for some feature \(i\) (in an input
\(x\)) to \num{0} if: (1) the saliency score for \(i\) is positive and
\(x_i=1\), or (2) the saliency score for \(i\) is negative and \(x_i=0\). This
prevents our version of the \jsma{} from selecting features that are already at
limits of valid feature values (\ie{} \num{1} and \num{0}). Moreover, we do not
select pixel pairs, as described in~\cite{papernot_limitations_2016}, as we
found our implementation to be at least as effective (often more) as the
original \jsma{}.

\shortsection{\dlrlossfull{}} The original formulation of \dlrloss{} requires
takes the ratio of the differences between: (1) the true logit and largest
non-true-class logit, and (2) the largest logit and the third largest logit. In
our evaluation, we used datasets that had less than three classes. For those
scenarios, we take the second largest logit.

\begin{table}
    \centering
    \resizebox{.49\textwidth}{!}{%
    \begin{tabular}{rccccccc}
        \toprule
         & \phishing & \nslkdd & \unswnb & \mnist & \fmnist   & \cifar{} & \malmem{}\\
        \midrule
        \multirow{2}{*}{Conv. Neurons}&-& -& -& (16,32)& (16,32) & (3,64,64,128,128,256,256, &-\\
        & & & & & &256,512,512,512,512,512,)& \\
        Kernel Size & -& -&- &3 &3 & 3&-\\
        Stride & -& -&- &1 &1 & 1&-\\
        Dropout Prob. & -&- & -& 0.4&0.4 & 0.5&-\\
        MaxPool Kernel &- & -& -&2 & 2& 2&- \\
        MaxPool Stride &- & -& -&2 & 2 & 2& -\\
        Linear Neurons& (15,)& (60,32)&(15,) & (128,)& (512,)&(512,) &(32,)\\
        Activation & ReLU &ReLU & ReLU& ReLU &ReLU & ReLU &ReLU\\
        Loss & CCE &CCE & CCE& CCE& CCE & CCE& CCE\\
        Optimizer & Adam&Adam &Adam & Adam&Adam & SGD&Adam\\
        Learning Rate & 1e-2&1e-2 & 1e-2&1e-3 &1e-3 &5e-2 &1e-2\\
        Epochs &40 &4 & 40&20 &20 & 300 &180\\
        Batch size & 32 & 128 &128  & 64 & 64& 128 & 64\\
        \bottomrule
    \end{tabular}}
    \caption{Hyperparameters}\label{tab:hyperparameters}
\end{table}
\begin{table}
    \centering
    \begin{minipage}[t]{0.45\textwidth}
    \resizebox{\textwidth}{!}{%
    \begin{tabular}{rccccccc}
        \toprule
         & \phishing & \nslkdd & \unswnb & \mnist & \fmnist & \cifar{} & \malmemfull{}  \\
        \midrule
        Attack & \pgd{} & \pgd{} & \pgd{} & \pgd{} & \pgd{} & \pgd{} & \pgd{}\\
        Epochs & 10 & 10 & 5 & 30 & 30 & 3 & 10\\
        \(\alpha\) & 0.01 & 0.01 & 0.01 & 0.01 & 0.01 &0.01 & 0.01\\
        Random Restart & 0.05 & 0.01 & 0.01 & 0.1 & 0.1 & 0.03 & 0.01\\
        \bottomrule
    \end{tabular}}
    \caption{Adversarial Training Hyperparameters}\label{tab:adv_hyperparameters}
    \end{minipage}
\end{table}

\begin{table}[!ht]
    \centering
    \resizebox{.28\textwidth}{!}{
    \begin{tabular}{ll}
        \toprule
        \textbf{\textit{Symbol}} & \textbf{\textit{Meaning}}\\
        \midrule
        \(x\)                    & original input\\
        \(x'\)                   & adversarial example\\
        \(\delta\)               & perturbation added to \(x\) \\
        \(w\)                    & \(x\) in \(\tanh\) space\\
        \(f\)                    & victim model\\
        \(f(x)\)                 & model logits\\
        \(c\)                    & number of classes \\
        \(\hat{y}\)              & true label\\
        \(y\)                    & softmax output\\
        \(k\)                    & closest class \\
        \(L\)                    & loss function\\
        \(\alpha\)               & single-step perturbation magnitude\\
        \(\epsilon\)             & total perturbation\\
        \(\jacobian\)            & Jacobian of a model\\
        \(\texttt{SM}\)          & Saliency Map\\
        \(p\)                    & parameter for some \lp{}-norm\\
        \(\theta\)               & time importance parameter\\
        \(B\)                    & budget equation\\
        \(b\)                    & budget value for a given equation\\
        \(\mathscr{A}\)          & space of attacks\\
        \bottomrule
    \end{tabular}}
    \caption{Symbol usage and meaning}\label{tab:symbols}
\end{table}

\ifarxiv
\else
\begin{table}
    \centering
    \resizebox{.3\textwidth}{!}{%
        \begin{tabular}{lllllll}
        \toprule
        \textbf{\fw[\#]} & \textbf{Opt.} & \textbf{CoV}& \textbf{RR} & \textbf{\lp{}} &
        \textbf{SM} &  \textbf{Loss} \\
        \midrule
        \texttt{JSMA} & \sgd{} & \textit{False} & \textit{False} & \lp[0] & \jsmasmap{} & \identityloss{}\\\texttt{DF} & \sgd{} & \textit{False} & \textit{False} & \lp[2] & \dfsmap{} & \identityloss{}\\\texttt{BIM} & \sgd{} & \textit{False} & \textit{False} & \lp[\infty] & \identitysmap{} & \crossentropy{}\\37 & \sgd{} & \textit{True} & \textit{False} & \lp[2] & \identitysmap{} & \identityloss{}\\\texttt{PGD} & \sgd{} & \textit{False} & \textit{True} & \lp[\infty] & \identitysmap{} & \crossentropy{}\\
171 & \adam{} & \textit{False} & \textit{False} & \lp[0] & \identitysmap{} & \dlrloss{}\\
191 & \adam{} & \textit{True} & \textit{False} & \lp[\infty] & \identitysmap{} & \crossentropy{}\\
\texttt{CW} & \adam{} & \textit{True} & \textit{False} & \lp[2] & \identitysmap{} & \cwloss{}\\
246 & \adam{} & \textit{False} & \textit{True} & \lp[0] & \jsmasmap{} & \dlrloss{}\\
\fab{} & \bwsgd{} & \textit{False} & \textit{False} & \lp[2] & \dfsmap{} & \identityloss{}\\
449 & \mbs{} & \textit{False} & \textit{False} & \lp[\infty] & \dfsmap{} & \crossentropy{}\\
460 & \mbs{} & \textit{False} & \textit{False} & \lp[2] & \identitysmap{} & \dlrloss{}\\
\apgdce{} & \mbs{} & \textit{False} & \textit{True} & \lp[\infty] & \identitysmap{} & \crossentropy{}\\
\apgddlr{} & \mbs{} & \textit{False} & \textit{True} & \lp[\infty] & \identitysmap{} & \dlrloss{}\\
        \bottomrule
        \end{tabular}
    }
    \caption{Attack Name Encodings.}\label{tab:translated}
\end{table}
\fi

\end{document}